\documentclass[10pt, a4paper]{extarticle}
 	
\usepackage{times}
\usepackage{fullpage}
\usepackage{authblk}
\usepackage{fancyhdr}							
\usepackage{amsfonts} 						
\usepackage{amssymb} 							
\usepackage[intlimits]{amsmath}		
\usepackage{graphicx} 						
\usepackage{subfigure} 						
\usepackage[numbers,square,comma,sort&compress]{natbib}				
\usepackage{units}								
\usepackage{textgreek}						
\usepackage[figuresright]{rotating}	
\usepackage[margin=17pt, font=small ,labelfont=bf]{caption} 			
\usepackage{icomma} 							
\usepackage{placeins}							
\usepackage{color}							
\usepackage{bm}	
\usepackage{lineno}
\usepackage{hyperref}							
\usepackage{amsbsy}
\usepackage{siunitx}		
					
\date{}

\newenvironment{addendum}{%
    \setlength{\parindent}{0in}%
    \begin{list}{Acknowledgement}{%
        \setlength{\leftmargin}{0in}%
        \setlength{\listparindent}{0in}%
        \setlength{\labelsep}{0em}%
        \setlength{\labelwidth}{0in}%
        \setlength{\itemsep}{12pt}%
        }
    }
    {\end{list}\normalsize}

\title{\textbf{Anatomizing deformation mechanisms \\in nanocrystalline Pd$_{90}$Au$_{10}$}}

\author[1]{Manuel~Grewer}
\author[1]{Christian~Braun}
\author[1]{Michael~J.~Deckarm}
\author[2]{Jochen~Lohmiller}
\author[2]{Patric~A.~Gruber}
\author[3]{Veijo~Honkim\"{a}ki}
\author[1]{Rainer~Birringer}

\affil[1]{Experimentalphysik, Universit\"at des Saarlandes, Saarbr\"ucken, Germany}
\affil[2]{Institute for Applied Materials, Karlsruhe Institute of Technology, Karlsruhe, Germany}
\affil[3]{Materials Science Group, European Synchrotron Radiation Facility, Grenoble, France}
\newcommand{\Dvol}{\langle D\rangle_{\mathrm{vol}}}
\modulolinenumbers[5]

\begin{document}

\maketitle


\noindent\textbf{We utilized synchrotron-based in-situ diffraction and dominant shear deformation to identify, dissect, and quantify the relevant deformation mechanisms in nanocrystalline $\mathrm{Pd}_{90}\mathrm{Au}_{10}$ in the limiting case of grain sizes at or below $\unit[10]{nm}$. We could identify lattice and grain boundary elasticity, shear shuffling operating in the core region of grain boundaries, stress driven grain boundary migration, and dislocation shear along lattice planes to contribute, however, with significantly different and nontrivial stress-dependent shares to overall deformation. Regarding lattice elasticity, we find that Hookean linear elasticity prevailed up to the maximal stress value of $\approx \unit[1.6]{GPa}$. Shear shuffling that propagates strain at/along grain boundaries increases progressively with increasing load to carry about two thirds of the overall strain in the regime of macroplasticity. Stress driven grain boundary migration requires overcoming a threshold stress slightly below the yield stress of $\approx \unit[1.4]{GPa}$ and contributes a share of $\approx\unit[10]{\%}$ to overall strain. Appreciable dislocation activity begins at a stress value of $\approx \unit[0.9]{GPa}$ to then increase and eventually propagate a maximal share of $\approx\unit[15]{\%}$ to overall strain. {In the stress regime below $\unit[0.9]{GPa}$, which is characterized by a markedly decreasing tangent modulus, shear shuffling and lattice- and grain boundary elasticity operate exclusively.} The material response in this regime seems indicative of nonlinear viscous behavior rather than being correlated with work- or strain hardening as observed in conventional fcc metals.
}

\section{Introduction}
When substantially reducing the crystallite (grain) size $D$ of polycrystalline metals into the nanometer regime, they show superior strength and hardness and are therefore potential candidate materials for a variety of applications. Over the last decade, this has motivated intense research efforts with the aim of gaining insight into the physics of grain size-dependent deformation mechanisms of nanocrystalline (NC) metals. Since the volume fraction of grain boundaries (GBs) scales with the reciprocal grain size, the abundance of GBs at the nanometer scale supplies barriers for intragranular slip and the nanometer-sized grains entail a reduced capacity of dislocation generation and interaction even at the upper limit of the nanometer scale of $\approx \unit[100]{nm}$. As a consequence, higher strength, lower activation volume, and higher strain-rate sensitivity have been observed \cite{Meyers2006,Dao2007}. With decreasing the grain size to the lower end of the nanoscale ($D \approx \unit[10]{nm}$ or below), it is expected that intragranular crystal plasticity becomes gradually or even dominantly replaced by GB-mediated deformation processes.

Simulations and experiments unraveled a variety of modes of plastic deformation related to GBs. So far the following processes have been identified: GB slip and sliding \cite{Vo2008,VanSwygenhoven2001,Weissmuller2011}, stress-driven GB migration (SDGBM) coupled to shear deformation and grain rotation \cite{Cahn2006,Cahn2004,Legros2008}, as well as shear transformation (ST) mediated plasticity \cite{Lund2005,Argon2006}, usually observed in glasses or colloidal systems \cite{Argon1979}, but here operating in the confined space related to the core region of GBs. Moreover, GBs and triple junction lines, locations where typically three GBs meet, may act as stress concentrators \cite{Gu2011,Asaro2005}, so effectively reducing the barrier for partial dislocation nucleation and emission. One of the intriguing aspects is that plastic deformation of NC metals requires that this variety of strain carriers interact in a synergistic way to make deformation happen in a compatible manner.

{In a recent work Ames et al. \cite{Ames2012} demonstrated that NC Pd$_{90}$Au$_{10}$ ($D \leq \unit[10]{nm}$) exhibits a pronounced strain-rate dependence prevailing up to large strains. Utilizing activation parameters to identify active deformation mechanisms, they found the measured activation volume and strain-rate sensitivity of NC Pd$_{90}$Au$_{10}$  to be compatible within the error margins with partial dislocation activity (PDA), STs (the generic flow event in metallic glasses) and SDGBM. Discrimination between just possible and relevant deformation mechanisms yet requires additional information. In a follow-up study, Grewer et al. \cite{Grewer2014} analyzed the activation energy of deformation in NC Pd$_{90}$Au$_{10}$ ($D \leq \unit[10]{nm}$) and studied SDGBM in detail. {They concluded} that ST-based shear shuffling operating in GBs dominates the overall deformation behavior. More fundamentally, they found that the activation energy $\Delta G$ exhibits a barrier height scaling $\Delta G \propto \tau^{3/2}$ which bears great similarity with the Johnson-Samwer \cite{Johnson2005} universal scaling law of yielding in metallic glasses; $ \tau$ is a measure of the applied load here. Finally, Skrotzki et al. \cite{Skrotzki2013} studied texture formation in NC Pd$_{90}$Au$_{10}$ induced by high-pressure torsion to find absence of any texture formation up to applied strains of $\gamma \approx 1$. Moreover, they observed that twinning and stacking fault formation is basically missing in this strain regime. Consequently, dislocation activity seems to play a minor role at the low end of the nanoscale.}

{Because of the complex interplay of disparate mechanisms, operating either sequentially or simultaneously, it remains to be clarified which role they play in responding to the intrinsic stress field and quantified which share to overall strain propagation is carried by them. Finding answers to these issues is a challenge for experiment and theory, as well as computer simulations. The latter have contributed to an atomic-level understanding of how different plasticity mechanisms interact on the nanoscale \cite{VanSwygenhoven2006,Yamakov2004,Vo2008a,Wu2013}. Clearly, the deformation of NC materials involves a broad spectrum of time and length scales beginning with the atomistic details of local shear in GBs, dislocation nucleation, -motion and -interaction, likewise, slip transfer across GBs to end in crack propagation or shear banding at the system level. Therefore, direct comparison between computer simulations and deformation experiments or advanced microscopy, as well as atom probe tomography still remains a tough case. Not least since the atomic interaction potentials used in simulation scenarios may not fully cover the genuine nature of the materials.}

{We address the experimental approach here and discuss which insight we gained from synchrotron-based diffraction measurements to in-situ study the microstructural evolution during deformation of NC Pd$_{90}$Au$_{10}$ {($D \leq \unit[10]{nm}$)} specimens prepared in a specific sample geometry termed shear compression specimen (SCS) (details are given in section 2). In addition, we analyzed dark field transmission electron micrographs at different strain values to extract possible changes of the grain size distribution function caused by SDGBM. The paper is organized as follows: first, the experimental setup and measurement procedures are introduced. Secondly, we discuss how individual XRD peak parameters have been extracted from the in-situ experiment and relate them to microstructural changes as well as associated deformation mechanisms. We will use this information in conjunction with the data from mechanical testing to quantify the relative deformation share of each identified process as a function of applied strain. Eventually, we display the shares of mechanisms to overall deformation as a function of applied load to unravel their stress-dependent {sequence} and/or simultaneous onset and persistence as well as revealing possible synergistic effects associated with their simultaneous presence.}
%
\begin{figure*}[t]
	\centering
		\includegraphics[width=0.95\textwidth]{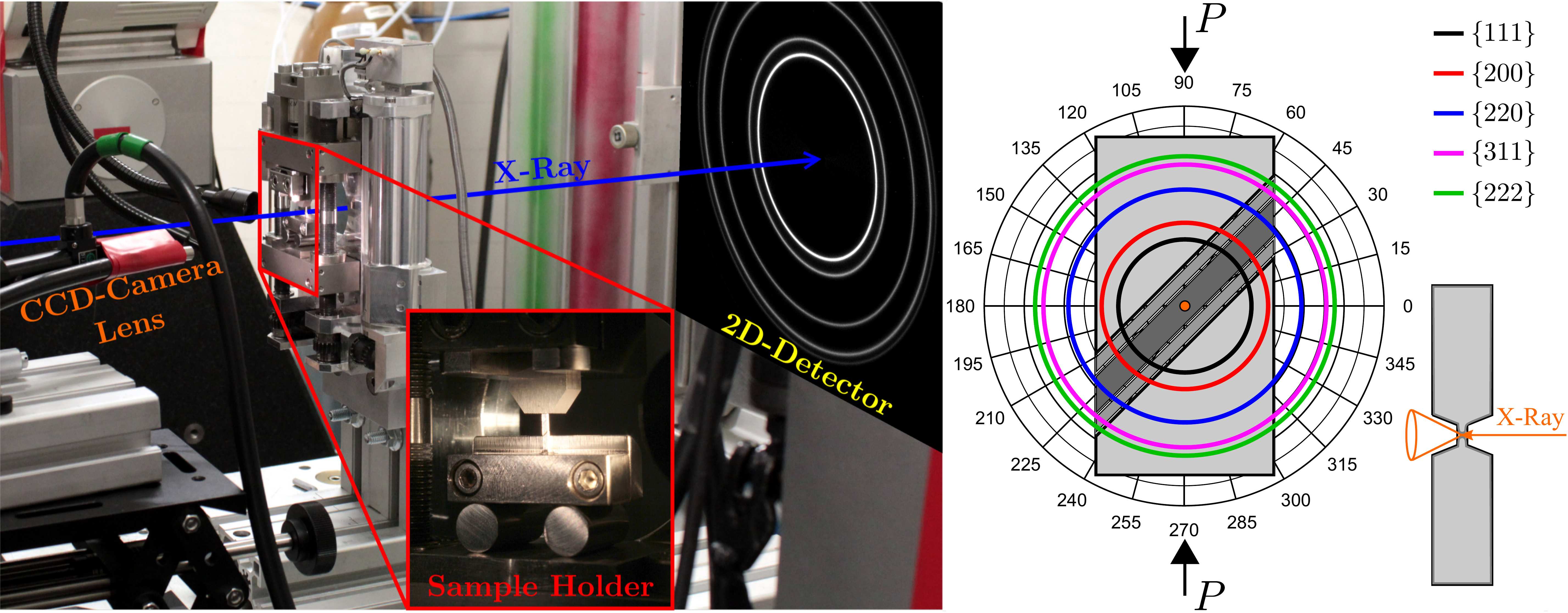}
	\caption{Photograph of the measurement setup at beamline ID15A at ESRF with the mechanical testing device mounted on top of a six axis goniometer. The inset shows a magnification of the compression stage with the roll bearing sample holder and a SCS in place. The sketch on the right displays a SCS in front and side view and serves as a reference frame for the polar plots shown throughout this document; the polar angle $\phi$ is measured counterclockwise and the three o'clock position defines $\phi=0^\circ$. In all cases the slit of the SCS is positioned $45^\circ$ from lower left to upper right and load $P$ is applied along the vertical direction ($\phi=90^\circ , 270^\circ$). Furthermore, all graphs showing \{hkl\}-data will refer to the same color code as used for the Debye-Scherrer rings in this sketch. The smaller side view shows the trapezoid slit geometry enabling undisturbed propagation of the diffracted beam.}
	\label{fig:ExpSetup}
\end{figure*}
%
 
\section{Material and methods}\label{Material-and-Methods}
Regarding the material system, we prepared NC Pd$_{90}$Au$_{10}$ alloys by inert-gas condensation (IGC) and compaction \cite{Birringer1989}, allowing {to explore} grain sizes at or even below $\unit[10]{nm}$. {IGC-prepared NC materials are statistically isotropic and homogeneous, characterized by a lognormal grain size distribution, and feature a GB-misorientation distribution that bears great similarity with the Mackenzie-distribution \cite{Birringer2003,Krill1998,Skrotzki2013}. As such, they manifest a model system of a polycrystal with random texture. Moreover, the PdAu alloy system is a fully miscible solid solution with a relatively high stacking fault energy $\approx \unit[180]{mJ/m^2}$ \cite{Schafer2011,Jin2011} in the Pd-rich alloys. Applying a compaction pressure of \unit[2]{GPa}, we produced disk-shaped samples (as-prepared samples) with a diameter of \unit[8]{mm} and a thickness of {$\unit[\approx 0.77]{mm}$}. The volume-weighted average grain-diameter $\Dvol$ of the as-prepared samples was determined from the Bragg peak broadening of X-ray diffraction (XRD) measurements using Klug \& Alexander's modified Williamson-Hall technique \cite{Klug1974,Markmann2008}. The diffraction experiments were performed on a laboratory diffractometer (PANalytical XPert Pro) operated in Bragg-Brentano focusing geometry and $\theta\,$--$\,\theta$ mode. The composition of as-prepared specimens was determined by EDX in a SEM (JEOL F 7000).}

{Mechanical testing was carried out using small-scale-specimens having {the} specific geometry shown in Fig. \ref{fig:ExpSetup}. Such specimens are termed shear compression specimens (SCS) \cite{Rittel2002} and were fabricated by employing spark erosion processing to as-prepared samples. The two parallel and oblique slits recessed on opposing faces form the gauge section. The dimensions of the SCS were $\unit[7]{mm} \times \unit[0.95]{mm} \times \unit[0.77]{mm}$ (H$\times$W$\times$T) and the gauge section was thinned to a thickness of $\unit[123]{\mu m}$. Applying load to the top and bottom faces forces the SCS to shear along the gauge section, which has been oriented $\unit[45]{^\circ}$ relative to the load direction. Such miniaturized SCS have been successfully applied to large strain testing of NC PdAu alloys by Ames et al. \cite{Ames2012}. Since plastic deformation is confined to the gauge section \cite{Ames2010}, the SCS is ideally suited for synchrotron based transmission experiments with the scattering vector probing the in-plane evolution of XRD peak parameters during deformation. To avoid shadowing of the scattering cones by the edges of the gauge slits, we deviate from the SCS design in \cite{Ames2012} by using a trapezoid slit profile (see Fig. \ref{fig:ExpSetup}) to adapt the geometry of the SCS to the diffraction measurement.}

Figure \ref{fig:ExpSetup} shows the experimental setup at the high energy microdiffraction endstation (HEMD) of beamline ID15A at the European Synchrotron Radiation Facility (ESRF). Mechanical testing was performed by a Kammrath \& Weiss tension/compression device made up of two load plungers moving symmetrically towards the center, thus keeping the gauge section in fixed position relative to the incoming beam. The lower plunger was replaced by a roll bearing wagon to substantially reduce the friction coefficient of the horizontal sample movement that goes along with the enforced shear deformation \cite{Ames2010}. The testing device with the installed SCS was mounted on top of a six axis goniometer, which enables exact placement of the sample with respect to the incoming synchrotron beam. The focused beam with a cross section of $\unit[8]{\mu m} \times \unit[20]{\mu m}$ was directed to the center of the thin gauge section, thus penetrating approximately 20 billion grains.The high energy X-ray radiation had a wavelength of $\lambda = \SI{0.178}{\angstrom}$ entailing an in-plane orientation (perpendicular to the incoming beam direction) of the scattering vector. The diffracted signal was recorded on a 2D-area detector (mar CCD), capable of recording one scan every seven seconds. The detector-sample distance was set to capture the first five Bragg reflections (Debye-Scherrer rings), so enabling a reasonable compromise between recording sufficiently high diffraction orders while still maintaining a good angular resolution. For data analysis, the 2D-scans were averaged over $2^{\circ}$-wide polar segments to give 180 radial line scans which where then fitted by Split-Pearson-VII functions to obtain peak parameters like $2\theta$ peak position, integral intensity and integral peak width. For more details related to data reduction, we refer to the recent work by Lohmiller et al. \cite{Lohmiller2014}.

{Sample deformation was recorded by a CCD-camera using an inline microscope lens for magnification and several light sources to illuminate the sample. The images were processed by digital image correlation (DIC) to extract true sample displacements to establish load-displacement curves {\cite{Grewer2013}}. As the gauge section of the SCS deforms under dominant shear and some superimposed pressure, there are no simple formulas to compute stress-strain curves from load-displacement data. Instead, we employed {finite element analyis (FEA)} using Abaqus, adapted to each individual sample geometry, to first generate load-displacement curves that best approximated the experimental load-displacement curves. Based on these fits, Von Mises equivalent stress-strain curves were then {extracted from the} FEA. More details on SCS-testing are given in \cite{Ames2012,Ames2010}. When placing the SCS samples in the sample holder, it needs some load until the SCS fully adapts to the frame of the testing machine. Such mounting problems generate uncertainties in the initial stage of the load displacement curves. To circumvent these uncertainties in the stress strain domain, we decided to extrapolate back from the clean linear stress strain regime to zero stress and strain.} {Therefore, this extrapolated regime (below about $0.3 \%$ strain) was not used for further analysis.}

{We also performed ex-situ TEM measurements on the deformed SCS to further characterize the evolution of the microstructure. Using focused ion beam (FIB) milling, thin lamellae were cut from the only elastically deformed bulk of the specimen as well as from the gauge section having experienced $\approx\unit[32]{\%}$ applied strain. The former serves as a reference for the undeformed material. The plastically deformed gauge lamella was oriented with its normal pointing perpendicular to the incoming X-ray beam direction. The grain microstructure of the lamellae were analyzed by an FEI Titan TEM, operated in dark-field mode by using different portions of the $\{$111$\}$-diffraction ring. Multiple areas were analyzed on each lamellae. As standard contrast thresholding approaches for automated grain analysis do not work well on micrographs of NC samples, the images were processed manually in ImageJ \footnote{rsbweb.nih.gov/ij/}. For each visible grain the longest and shortest chord length was determined using the line measurement tool in ImageJ. A consistency check, involving the repeated measurement of an image set containing ca. 200 grains, showed good reproducibility of individual grain diameters within \unit[0.5]{nm} uncertainty. To improve statistics and reduce the overall error, at least 1000 grains were counted per lamella (deformed and undeformed).}

%
\begin{figure}[ht]
	\centering
		\includegraphics[width=0.45\textwidth]{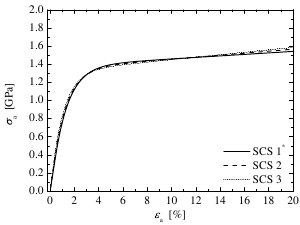}
	\caption{Stress-strain curves of three individual Pd$_{90}$Au$_{10}$-SCS deformed at $\dot{\varepsilon}_\mathrm{a}\approx\unit[1.5\cdot 10^{-3}]{1/s}$. The $^*$ marks the sample that will be analyzed and discussed throughout this document.}
	\label{fig:StressStrain}
\end{figure}
%

\section{Results and discussion}
\subsection{{Mechanical testing}}
{Figure \ref{fig:StressStrain} displays stress-strain curves of three Pd$_{90}$Au$_{10}$-SCS, deformed at a constant machine speed of $\unit[0.15]{\mu m/s}$ which translates into applied strain rates $\dot{\varepsilon}_\mathrm{a}$ between $\unit[1.1\cdot 10^{-3}]{1/s}$ and $\unit[1.8\cdot 10^{-3}]{1/s}$. Despite the slightly different applied strain rates (owing to small deviations in sample geometry), these curves not only demonstrate good reproducibility but also suggest that generic intrinsic material behavior is reflected in the stress-strain response rather than influence of extrinsic effects. This is also true for the scattering data collected during deformation. In what follows, we will therefore exemplarily discuss the diffraction results of the SCS marked by $^*$ in Fig. \ref{fig:StressStrain}.}

\subsection{Bragg peak position and lattice strain}
\begin{figure*}[b!]
     \subfigure[]{\includegraphics[width=0.48\textwidth]{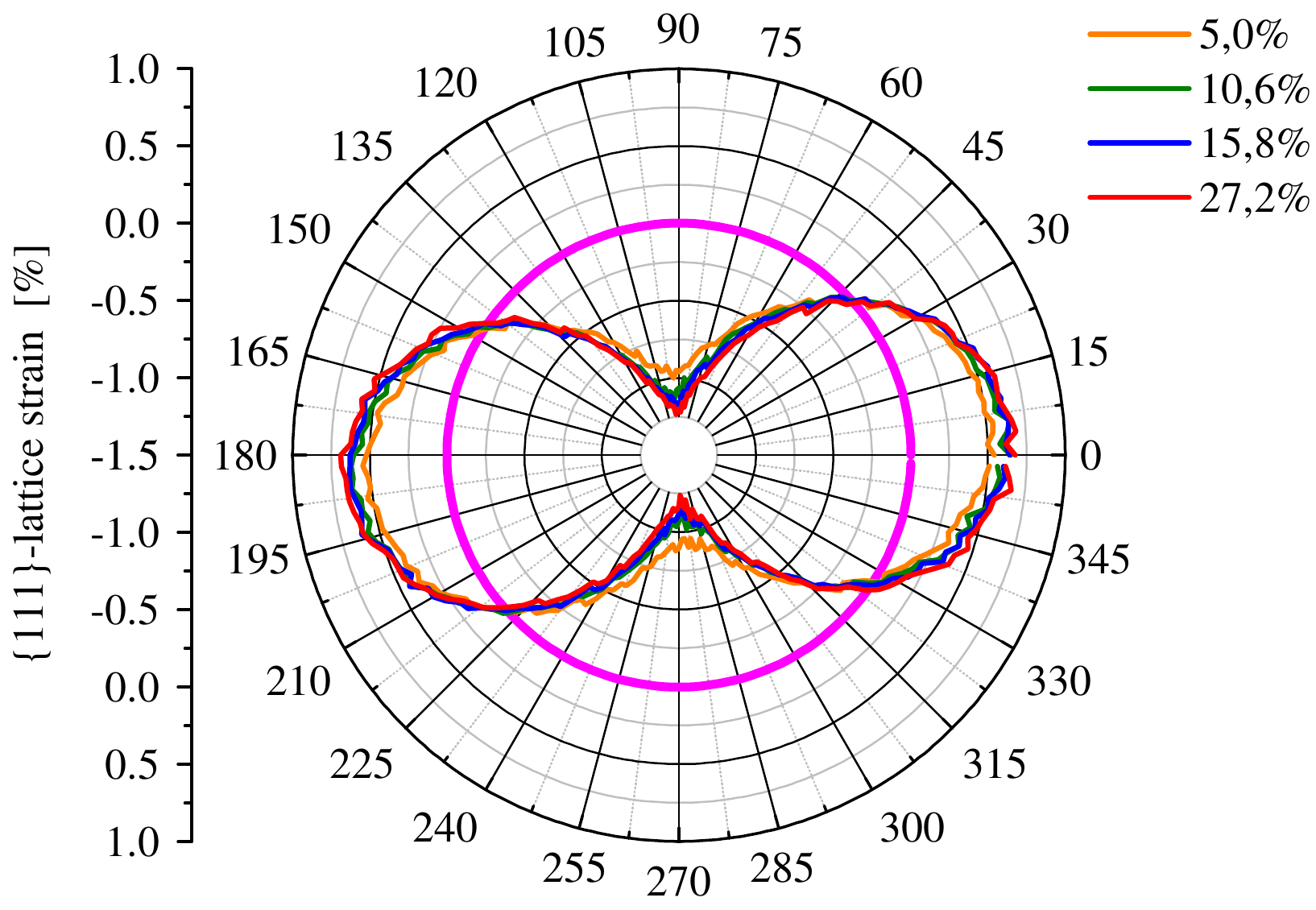}\label{fig:LStrainPolar}}
 \hspace*{.3cm}
     \subfigure[]{\includegraphics[width=0.45\textwidth]{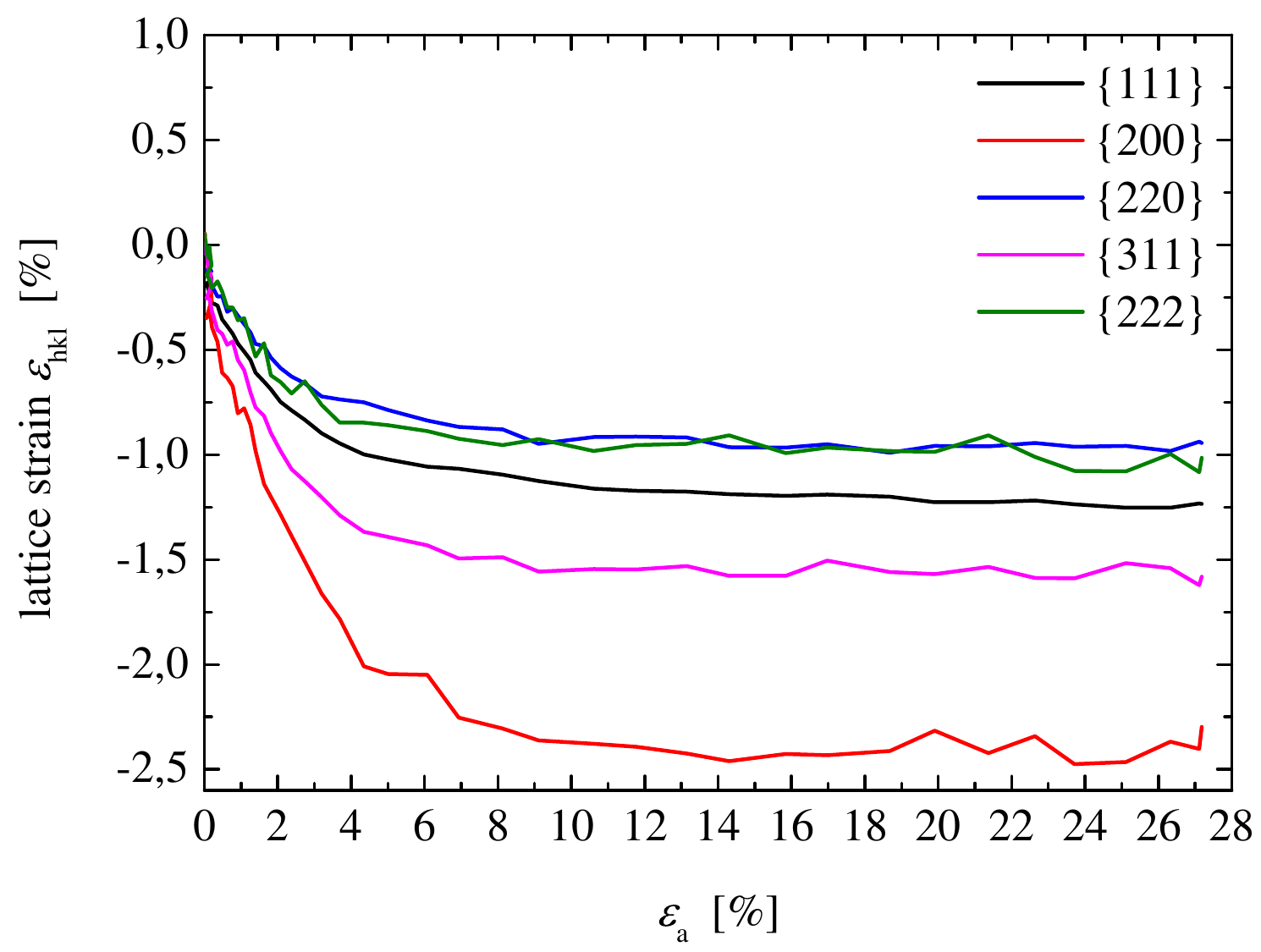}\label{fig:ElasticAnisotropy}}
 \caption[]{(a) {Polar plot of \{111\}-lattice strains of four selected deformation states as indicated in the figure legend. The magenta colored circle marks zero lattice strain.} (b) Compressive lattice strains of different \{hkl\}-lattice planes as a function of applied strain $\varepsilon_\mathrm{a}$ evaluated along the load direction ($\phi=90^\circ , 270^\circ$).}
 \label{fig:LStrain}
\end{figure*}

First, we look at the $2\theta$-angles characterizing the different \{hkl\}-Bragg peak positions. Using Bragg's law, it is straightforward to compute the corresponding lattice spacing $d_\mathrm{hkl}=\lambda(2\sin\theta_\mathrm{hkl})^{-1}$, where $\theta_\mathrm{hkl}$ is half of the diffraction angle and $\lambda = \unit[0.178]{\AA}$ is the wavelength of the synchrotron radiation. During deformation the lattice spacings are expected to change so that the resulting elastic lattice strain $\varepsilon_\mathrm{hkl}$ can be extracted from
\begin{equation}
  \varepsilon_\mathrm{hkl} = \frac{\Delta d_\mathrm{hkl}}{d_\mathrm{hkl,0}} = \frac{\sin\theta_\mathrm{hkl,0}}{\sin\theta_\mathrm{hkl}}-1, \label{eq:LStrain}
\end{equation}
where $\Delta d_\mathrm{hkl}=d_\mathrm{hkl}-d_\mathrm{hkl,0}$ and the index $_0$ corresponds to values of the initially undeformed sample. This definition implies positive lattice strain with increasing lattice spacing (tension), while a decrease (compression) leads to negative strain values. A polar plot of values of the \{111\}-lattice strain of a NC Pd$_{90}$Au$_{10}$-SCS is displayed in Figure \ref{fig:LStrainPolar} for several selected deformation states.
%
%
As the scattering vector is oriented perpendicular to the diffracting lattice planes, the signal along the load direction corresponds to a stack of horizontally oriented \{111\}-lattice planes and likewise for higher diffraction orders. Interestingly, the maximal strain values are not observed exactly along the load direction ($\phi=90^\circ, 270^\circ$) but at $\phi=95^\circ, 275^\circ$ instead. In fact, the whole graph appears to be rotated $\approx 5^\circ$ counterclockwise, indicating a transformation of the external load state, which, as can be verified by FEA, is related to the specific sample geometry (see sketch in Fig. \ref{fig:ExpSetup}).

\begin{figure*}[t]
     \subfigure[]{\includegraphics[width=0.46\textwidth]{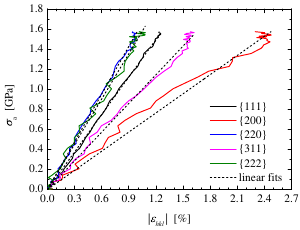}\label{fig:Hook}}
 \hspace*{.3cm}
     \subfigure[]{\includegraphics[width=0.425\textwidth]{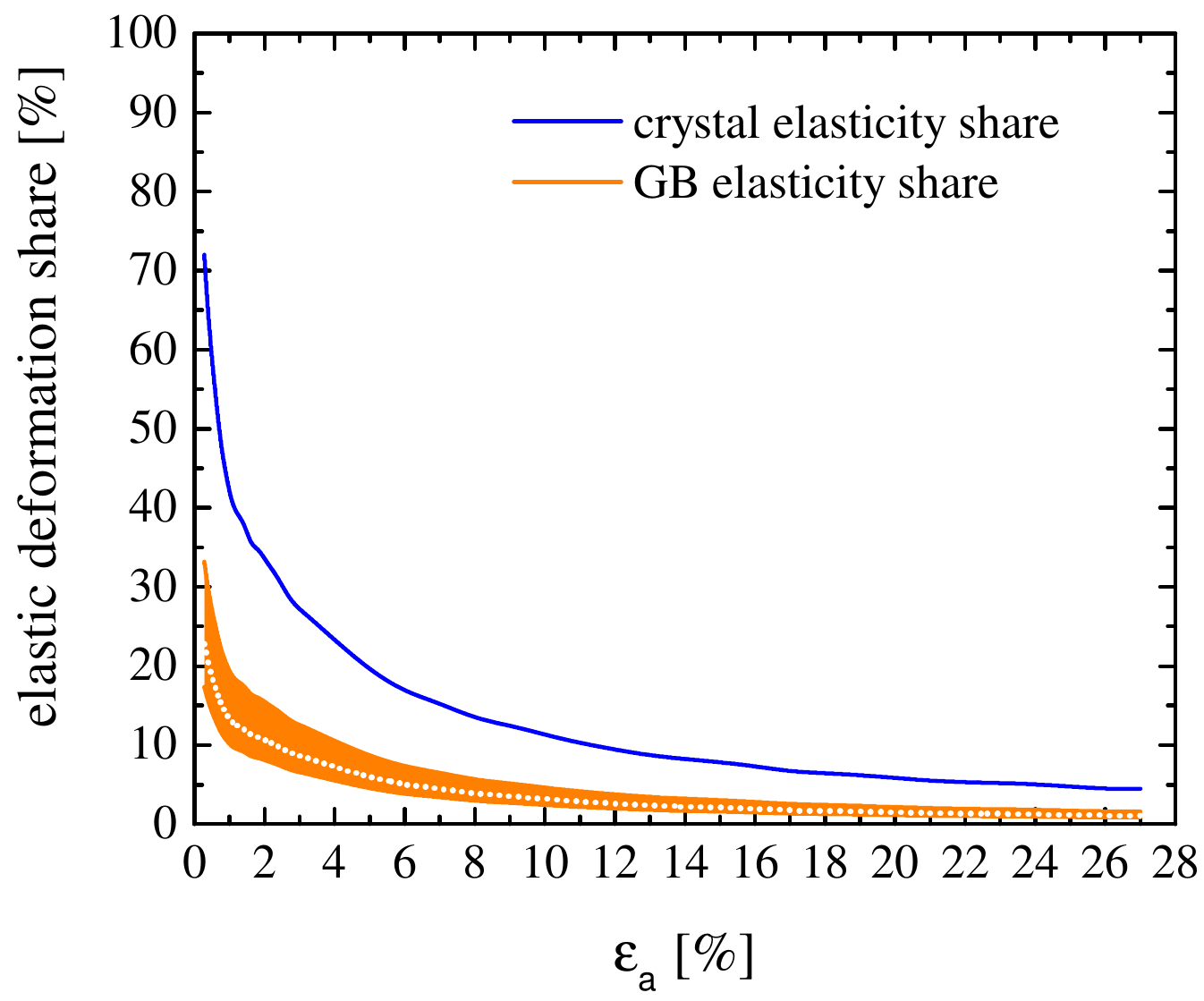}\label{fig:ElasticShares}}
 \caption[]{(a) Applied stress $\sigma_\mathrm{a}$ as a function of the absolute value of lattice strain $|\varepsilon_\mathrm{hkl}|$ along the load direction. Linear fits show that the elastic lattice deformation essentially follows Hook's law up to large stresses. Only the most compliant \{200\}-planes deviate from straight line behavior, indicating possible nonlinear elastic contributions. {(b) Share of crystal lattice elasticity to overall deformation (blue). The orange ribbon represents the spread between the upper and lower bound values of the share of GB elasticity to overall deformation (see text for details). The dotted line indicates the share corresponding to the mean Young's modulus of upper und lower bound.}}
\label{fig:HookAndShares}
\end{figure*}

Clearly, increasing load and hence applied strain correlate with increasing lattice strain, either compressive or tensile. In Fig. \ref{fig:ElasticAnisotropy}, we show the compressive lattice strain $\varepsilon_\mathrm{hkl}$ of all five measured Bragg reflections as a function of applied strain $\varepsilon_\mathrm{a}$, evaluated along the load direction ($\phi=90^\circ$, $\phi=270^\circ$). Due to the elastic anisotropy of PdAu, lattice planes with different Miller indexes experience different magnitudes of lattice strain and, typical for face centered cubic (fcc) metals, the \{200\}-planes are elastically most compliant. Apart from the different magnitudes, the lattice strain development of all planes follows roughly the shape of the stress-strain curve. This behavior is not unexpected though, since the evolution of lattice strain essentially describes the elastic deformation of the myriad of nanometer-sized crystal lattices. Interestingly, when plotting the applied stress as a function of the absolute value of lattice strain $|\varepsilon_\mathrm{hkl}|$ along the load direction, as shown in Fig. \ref{fig:Hook}, a linear correlation between stress and strain is observed. This evidence implies that even at such high stress values ($\sigma_\mathrm{a} > \unit[1.5]{GPa}$) nonlinear elastic behavior is absent but Hooke's law $\sigma_\mathrm{a}=C_\mathrm{hkl}\,\varepsilon_\mathrm{hkl}$, where $C_\mathrm{hkl}$ is the \{hkl\}-dependent elastic modulus, still applies. The only exception are the most compliant \{200\}-planes that slightly deviate from the straight line in Fig. \ref{fig:Hook}, suggesting presence of nonlinear elastic contributions. {Nevertheless, evaluation of the arithmetic mean of the slopes of the linear fits $C_\mathrm{hkl}$, weighted by their lattice plane multiplicity factor, results in an effective Young's modulus $\langle C\rangle = \unit[118\pm 4]{GPa}$, that is remarkably close to the Young's modulus of coarse grained Pd$_{90}$Au$_{10}$, which was determined by ultrasound to give a value of $\unit[127 \pm 5]{GPa}$ \cite{Leibner2015}.} {We therefore think that the implicit assumption of a homogeneous mean stress equal to the applied stress is valid.}
%
%

{We now exploit the contributions of crystal- and GB elasticity to overall deformation. In mechanical equilibrium, the generalized capillary equation \cite{Weissmuller1997} prescribes that GB stresses are balanced by bulk stresses of the abutting crystallites. Therefore, we assume that crystal elasticity goes along with elastic behavior of GBs, which however have significantly reduced elastic moduli \cite{Grewer2011}. The total elastic strain, $\varepsilon_\mathrm{elastic}$, related to the sample length measured along the loading direction, is then given by $\varepsilon_\mathrm{elastic} = (1- \chi) \varepsilon_\mathrm{x} + \chi \varepsilon_\mathrm{GB}$, where $\varepsilon_\mathrm{x}$ is the elastic strain due to crystal elasticity, and $\varepsilon_\mathrm{GB}$ is the strain contribution of the myriads of GBs occupying a length share $\chi$. Consequently, $(1- \chi)$ represents the length share of crystallites along the loading direction. The parameter $\chi$ is directly connected with the grain size $ \Dvol$ and can be computed {if} the width $\sigma$ of the grain size distribution and the structural width of the GBs, $\delta$, {are} known \cite{Grewer2011,Leibner2015}}. 

{The value of $\varepsilon_\mathrm{x}$ can be straightforwardly approximated by setting $ \varepsilon_\mathrm{x} = \langle\varepsilon_\mathrm{hkl}\rangle $, where the average elastic strain in loading direction $\langle\varepsilon_\mathrm{hkl}\rangle$ is represented by the arithmetic mean of the five curves displayed in Fig. \ref{fig:ElasticAnisotropy}, weighted by their respective X-ray lattice plane multiplicity factors. The GB elastic strain can not be extracted directly from diffraction data since the randomness of the interparticle interference generates a rather diffuse background intensity, which is in general not amenable to extract strain information. {Therefore, we write an effective Hooke's law for the aggregate of GBs $\sigma_{\mathrm{a}}=\langle E_{\mathrm{GB}}\rangle \, \varepsilon_{\mathrm{GB}}$, where $\langle E_{\mathrm{GB}}\rangle$ is the effective Young's modulus of GBs and for the aggregate of crystallites $\sigma_{\mathrm{a}}=\langle C \rangle \, \varepsilon_{\mathrm{x}}$.} Equating the right hand side of both Hooke's laws, we find the sought expression for $\varepsilon_\mathrm{GB} = \langle\varepsilon_\mathrm{hkl}\rangle (\langle C \rangle / \langle E_{\mathrm{GB}}\rangle)$.}

{The shares of the myriads of nanocrystals and GBs to overall strain $\varepsilon_\mathrm{a}$ can be deduced from
\begin{equation}
  \frac{\varepsilon_\mathrm{elastic}}{\varepsilon_\mathrm{a}} = (1- \chi) \frac{ \langle\varepsilon_\mathrm{hkl}\rangle }{\varepsilon_\mathrm{a}} + \chi \frac{\langle\varepsilon_\mathrm{hkl}\rangle }{\varepsilon_\mathrm{a}} \frac{\langle C \rangle}{\langle E_{\mathrm{GB}}\rangle}.  \label{eq:Shares}
\end{equation} 
The first term on the right hand side of Eq. \ref {eq:Shares} represents the crystal elastic share to overall deformation and the second term accounts for the GB share.} With $\sigma = 1.5$ (see Fig \ref{fig:GSinitial} and \ref{fig:GSdeformed}) and $\delta = \unit[0.76 \pm0.01]{nm}$ \cite{Leibner2015}, the evolution of the share of crystal elasticity as a function of increasing deformation can be computed and is shown in Fig. \ref{fig:ElasticShares}. To estimate the share of GB elasticity to overall deformation, we need to approximate the ratio $\langle C \rangle / \langle E_{\mathrm{GB}}\rangle$. To fix $\langle C \rangle $, we make use of the value $\langle C\rangle = \unit[118\pm 4]{GPa}$ deduced from Fig. \ref{fig:Hook} and note that this value represents the effective Youngs's modulus of the aggregate of crystallites, which must be slightly larger than the true Reuss lower bound of the aggregate. We recently determined the effective (high frequency) Young's modulus of GBs by means of ultrasound measurements on unloaded samples to $\unit[61 \pm8]{GPa}$ \cite{Leibner2015}. However, under an applied load, we cannot rule out, even at the onset of deformation, configurational changes to occur in the core region of GBs that should give rise to an increase of $\langle E_{\mathrm{GB}}\rangle$. Thus, the zero-load high frequency Young's modulus manifests a lower bound to the effective GB modulus $\langle E_{\mathrm{GB}}\rangle$. For the sake of argument, we assume $\langle E_{\mathrm{GB}}\rangle = \unit[61 \pm8]{GPa} $, consequently, the share of strain of GBs to overall deformation assumes upper bound character. In contrast, an ultimate upper bound for $\langle E_{\mathrm{GB}}\rangle$ entails $\langle E_{\mathrm{GB}}\rangle = \langle C \rangle $, and so results in a lower bound for the GB share to overall deformation. The spread of the discussed bounds is indicated by the orange ribbon in Fig. \ref{fig:ElasticShares}. In what follows, we consider the share of GB elasticity to overall deformation corresponding to the mean Young's modulus of the upper- and lower bound, represented by the dotted line in Fig. \ref{fig:ElasticShares}.

\subsection{Integral peak width: microstrain and grain size evolution}\label{Integral-peak-width}
In this paragraph we focus on the evolution of the integral peak width of the diffraction reflexes (Debye-Scherrer rings). The two main factors that contribute to the broadening of diffraction peaks of a NC material are the small size of the scattering domains, i.e. small grain size, and the so called microstrain $e$. The latter arises from small variations in lattice spacings of neighboring grains, which are commonly attributed to strain fields near GBs or lattice distortions caused by stored dislocations \cite{Ung'ar1999}. However, virtual diffraction experiments in conjunction with molecular dynamics (MD) based determination of atomic level strains \cite{Stukowski2009} have shown that the XRD microstrain is insensitive to strain fields that decay faster than $r^{-3/2}$, specifically, the exponentially decaying local strains near GBs. Instead, the microstrain broadening in NC materials in the limit of small grain sizes is dominated by long range displacement fields that extend throughout the grains and arise from compatibility constraints imposed on the grains to form a space filling tessellation of crystalline objects. In fact, studying room-temperature grain growth in NC Pd ($D_{\mathrm{initial}} \approx \unit[10]{nm}$), Ames et al. \cite{Ames2008} have found the XRD microstrain to be inversely proportional to the average grain size $e \propto 1/\langle D\rangle$, suggesting that microstrain is in fact related to tesselation constraints and therefore is a generic contribution in any polycrystalline material. Due to the $1/\langle D\rangle$-scaling, microstrain becomes particularly pronounced at the nanoscale even in an undeformed, load-free NC structure.
\begin{figure*}[ht]
     \subfigure[]{\includegraphics[width=0.45\textwidth]{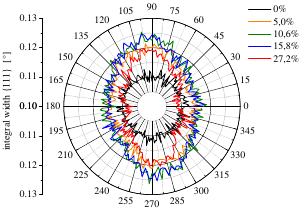}\label{fig:IntWidthPolar}}
 \hspace*{.3cm}
     \subfigure[]{\includegraphics[width=0.45\textwidth]{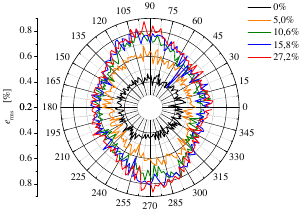}\label{fig:RMSMicrostrainPolar}}\\
     \subfigure[]{\includegraphics[width=0.45\textwidth]{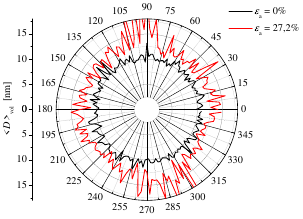}\label{fig:GrainSizeEvolutionPolar}}
 \hspace*{.05cm}
     \subfigure[]{\includegraphics[width=0.43\textwidth]{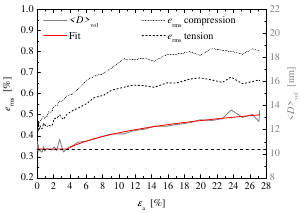}\label{fig:GrainSizeEvolution}}
 \caption[]{(a) Polar plot of the integral width of the \{111\}-reflection for several deformation states. (b) Polar plot of the root mean square microstrain $e_\mathrm{rms}$. (c) Polar plot of $\Dvol$ for the initially undeformed SCS and at $\varepsilon_\mathrm{a}=\unit[27.2]{\%}$. (d) Evolution of $\Dvol$ and $e_\mathrm{rms}$ as a function of applied strain. $\Dvol$-values have been averaged over all polar angles (solid line) while microstrain values are shown for $\phi=\unit[0]{^\circ}$ (dashed line), corresponding to lateral tension, and $\phi=\unit[90]{^\circ}$ (dotted line), equivalent to the load direction.}
 \label{fig:GrainGrowth}
\end{figure*}

Upon loading, the deduced microstrain also reflects the elastic anisotropy of the material, as grains with their hard axis being oriented parallel to the load direction will experience less compression than grains oriented along an elastically weaker axis. This behavior is readily illustrated by the polar plot in Fig. \ref{fig:IntWidthPolar}, where the integral \{111\}-peak width is increasing along the load direction with increasing strain up to $\varepsilon_\mathrm{a}=\unit[15.8]{\%}$. Further deformation to $\varepsilon_\mathrm{a}=\unit[27.2]{\%}$ leads to a decrease of the integral peak width, revealing a superimposed change of grain size. Effectively, {the} decreasing peak width is inversely proportional to the size of the scattering domains. {Thus, the decrease of integral peak width at large strains in Fig. \ref{fig:IntWidthPolar} can be attributed to an increase of the volume averaged grain size $\Dvol$.}

{To analyze the microstrain and grain size evolution, we make use of the single line method, introduced by de Keijser et 
al. \cite{Keijser1982}, which allows to extract separate values of grain size and microstrain for each \{hkl\}-reflection. The widely used Williamson-Hall approach led to poorly conditioned results, certainly due to the elastic anisotropy of PdAu but also related to the fact that the diffracted intensity has been recorded in transmission geometry, implying a slight variation of the {orientation of the} scattering vector with the diffraction angle which renders the Williamson-Hall approach intractable; further technical details can be found in \cite{Lohmiller2014}. Using the single line method, deconvolution of size and microstrain contributions results in the following equations for the volume averaged grain size $\Dvol$ and the microstrain $e$
\begin{eqnarray}
\Dvol^\mathrm{hkl} &=& \frac{\lambda}{\beta_\mathrm{L}\cos\theta_\mathrm{hkl}}\\
e^\mathrm{hkl} &=& \frac{\beta_\mathrm{G}}{4\tan\theta_\mathrm{hkl}},
\end{eqnarray}
where $\beta_\mathrm{L}$ and $\beta_\mathrm{G}$ are the Lorentzian and Gaussian broadening shares as given in \cite{Keijser1982}. Analysis of the diffraction data related to the in-situ deformation of the investigated Pd$_{90}$Au$_{10}$-SCS delivers microstrain and volume averaged grain size values for the five measured diffraction orders. We compute the arithmetic mean of the individual \{hkl\}-dependent values to obtain the microstrain and $\Dvol$ values displayed in Fig. \ref{fig:RMSMicrostrainPolar} and \ref{fig:GrainSizeEvolutionPolar}, respectively. For convenience, we have converted $e$ to the commonly used root mean square microstrain $e_\mathrm{rms}=(2/\pi)^{1/2}\,e$.}

{In the undeformed state, the microstrain amounts to roughly \unit[0.4]{\%} along all polar angles (Fig. \ref{fig:RMSMicrostrainPolar}). With increasing deformation, the microstrain steeply rises up to \unit[0.8]{\%} along the load direction and \unit[0.65]{\%} in the lateral tensile direction, where the relevant increase occurs at strains below $\varepsilon_\mathrm{a}=\unit[10.6]{\%}$ .} This becomes even more apparent in Fig. \ref{fig:GrainSizeEvolution}, where the evolutions of $e_\mathrm{rms}$ along the vertical compressive (dotted) and lateral tensile (dashed) direction are presented as a function of the applied strain. Interestingly, the elliptic shape of the microstrain polar plot corresponds well with the lattice strain development shown in Fig. \ref{fig:LStrainPolar}. The obvious connection between lattice strain $\varepsilon_\mathrm{hkl}$ and microstrain $e_\mathrm{rms}$ corroborates the notion that microstrain in NC metals is essentially a measure of long range displacement fields originating from compatibility constraints. We cannot discriminate between possible additional microstrain contributions resulting from stress concentrators at triple junctions and bending of lattice planes both caused by the likely presence of GB-mediated deformation modes \cite{Ovidko2007,Bachurin2010}. 

{Regarding grain size evolution, we first focus on the circular shape of the polar plot of the initial grain size (black) in Fig. \ref{fig:GrainSizeEvolutionPolar}. It indicates that the undeformed sample is made up of roughly equiaxed grains with a mean diameter of $\Dvol=\unit[10.4\pm 0.7]{nm}$. This value is in close agreement with the grain size $\Dvol^\mathrm{IGC}\approx\unit[9]{nm}$ of the as-prepared NC material, determined immediately after IGC by a Williamson-Hall analysis of XRD pattern recorded on an XPert Pro laboratory diffractometer in Bragg-Brentano focusing geometry, providing the required fixed orientation of the scattering vector. Therefore, we feel confident that the single line method is capable of catching the essential features of relative change of grain size.}

In Fig. \ref{fig:GrainSizeEvolution} the evolution of $\Dvol$, averaged over all polar angles, is displayed as a function of the applied strain. Remarkably, the grain size remains nearly constant at $\Dvol=\unit[10.4]{nm}$ up to $\varepsilon_\mathrm{a}\approx \unit[4]{\%}$. Beyond that point grain growth is triggered and gradually advances with increasing plastic strain to reach a grain size of $\Dvol=\unit[13.4 \pm 2.7]{nm}$ at $\varepsilon_\mathrm{a}=\unit[27.2]{\%}$. The fact that grain size is still growing, while the microstrain values shown in Fig. \ref{fig:GrainSizeEvolution} reach a plateau, is consistent with the decrease of the integral peak width in Fig. \ref{fig:IntWidthPolar} beyond $\varepsilon_\mathrm{a}\approx \unit[15.8]{\%}$. 

Comparing initial ($\Dvol=\unit[10.4 \pm 0.7]{nm}$) and final ($\unit[13.4 \pm 2.7]{nm}$) grain size values, we find that the deconvolution of microstrain and size broadening of pronouncedly stressed SCSs goes along with a higher error margin for the grain size values. Moreover, the deconvolution of size and strain seems severely affected by the high stress values along the vertical compressive direction (compare Fig. \ref{fig:GrainGrowth}(b) and (c)) leading to a fictitious enhancement of size along the loading direction; we come back to this issue later. Accepting sources and margin of errors, the polar plot of the deformed state appears approximately circular implying that a possible evolution of grain shape anisotropy (growth texture) plays, if at all, a minor role. The observation that stress generates approximately \textit{isotropic} growth seems to be indicative of stress driven grain boundary migration (SDGBM), synonymously termed coupled motion. Generally, it comprises the migration of a GB parallel and perpendicular to an applied shear stress $\tau$. The coupling factor $\beta=v_\parallel/v_\bot$ \cite{Cahn2006} describes the ratio of migration parallel ($v_\parallel$) and perpendicular ($v_\bot$) to the applied stress in terms of the velocities $v$ of the moving boundary. Since the NC samples prepared by IGC are statistically isotropic and homogeneous, we would expect that local migration anisotropies of individual boundaries should effectively average out. 

{For the sake of argument, let us assume that the vertical compressive stress generates  anisotropy by delivering the maximal driving force (shear stress) for $v_\bot$ of appropriately oriented GBs. We then expect that those GBs, having normals aligned towards $\phi \approx \unit[0]{^\circ}$ or $\unit[180]{^\circ}$, should exhibit maximal growth along this direction. A faint signature related to this rationale may be seen in Fig. \ref{fig:GrainSizeEvolutionPolar}. On the other hand, we may consequently argue that the smaller but effective lateral tension should then lead to comparatively less growth along the load direction ($\phi=90^\circ , 270^\circ$). However, this plausible reasoning is in contradiction to what has been extracted from experiment.} Therefore, we are confident that the growth indicated along the load direction is in fact fictitious. What is more, SDGBM has been seen also in NC Pd$_{90}$Au$_{10}$ after high pressure torsion deformation \cite{Skrotzki2013}. A nearly doubling of grain size has been deduced, also based on X-ray Bragg peak profile analysis, when the shear-strain approached $\gamma \approx 0.5$. But no sign of concomitant texture formation could be observed, thus lending independent support to the assertion of isotropic growth.
 
{Although it appears reasonable that SDGBM is the root cause of grain growth, we should not conceal that curvature-driven grain growth \cite{Ames2008} and dislocation glide could also lead to changes in $\Dvol$. However, the former process can be ruled out since all experiments were performed at room temperature (RT) and the microstructure of NC Pd$_{90}$Au$_{10}$ has been verified being stable (kinetically frozen) up to $\unit[60]{^\circ C}$. Loading the SCS and assuming availability of dislocations, unidirectional dislocation glide along the gauge section becomes favored and should manifest as grain shape anisotropy rather than isotropic grain growth. Specificially, assuming glide occurs on \{111\}-lattice planes, it is implied then that such planes which are already oriented along the gauge section or rotate with increasing load towards this direction would experience the largest Schmid factor and should therefore be the most active glide planes. As a result, the grain chord length measured along the glide-plane normal should shrink and grain elongation should appear along the gauge section. In particular, we would expect a decrease in grain size in the angular regime around $\unit[135]{^\circ}$, $\unit[315]{^\circ}$ and a concomitant increase towards {$\approx \unit[45]{^\circ}$}, $\unit[225]{^\circ}$. However, such a behavior is not corroborated by experiment (Fig. \ref{fig:GrainSizeEvolutionPolar}).}

{We might even argue that anisotropic growth related to both SDGBM and dislocation glide contributes in effect (due to their simultaneous presence) to apparent isotropic growth. However, as discussed above the leading orientation dependence of both effects is well within the resolution of experiment and, therefore, allows us to discriminate between simultaneous presence of both processes or contribution of a single process. Clearly, our experimental evidence reflecting that grain growth evolves basically grain-shape preserving favors SDGBM as the root cause of growth. Yet, this rationale relies upon the assumption that both processes contribute with well resolvable growth rates to overall growth. Therefore, to this end we cannot discard the presence of dislocation glide since its effect on grain-shape change may be so small that it is beyond the resolution associated with probing size by X-ray line broadening; we come back to this problem in the following section.}   
%
\begin{figure*}[b!]
\hspace*{1.2cm}
     \subfigure[]{\includegraphics[width=0.23\textwidth]{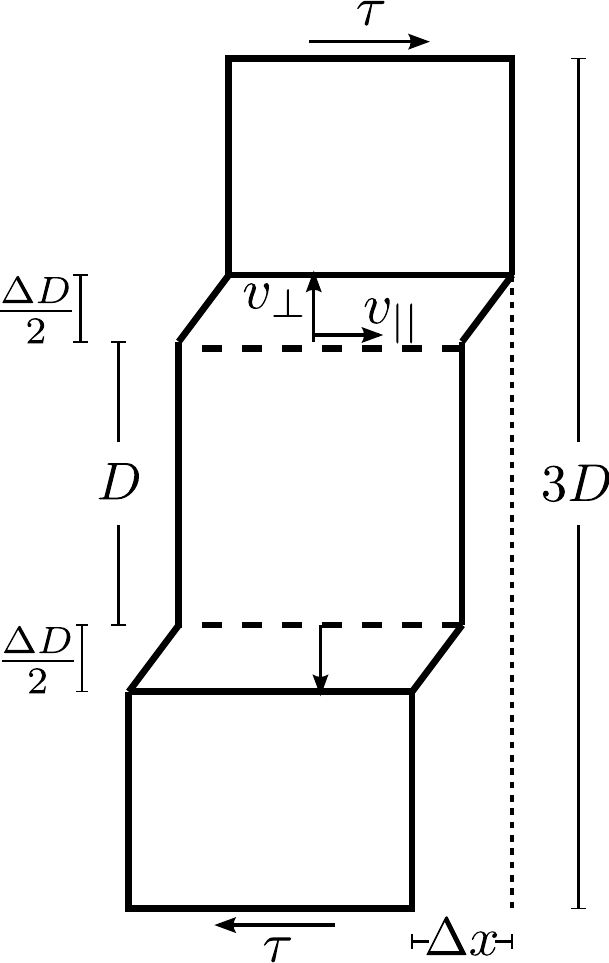}\label{fig:CouplingSketch}}
 \hspace*{1.5cm}
     \subfigure[]{\includegraphics[width=0.45\textwidth]{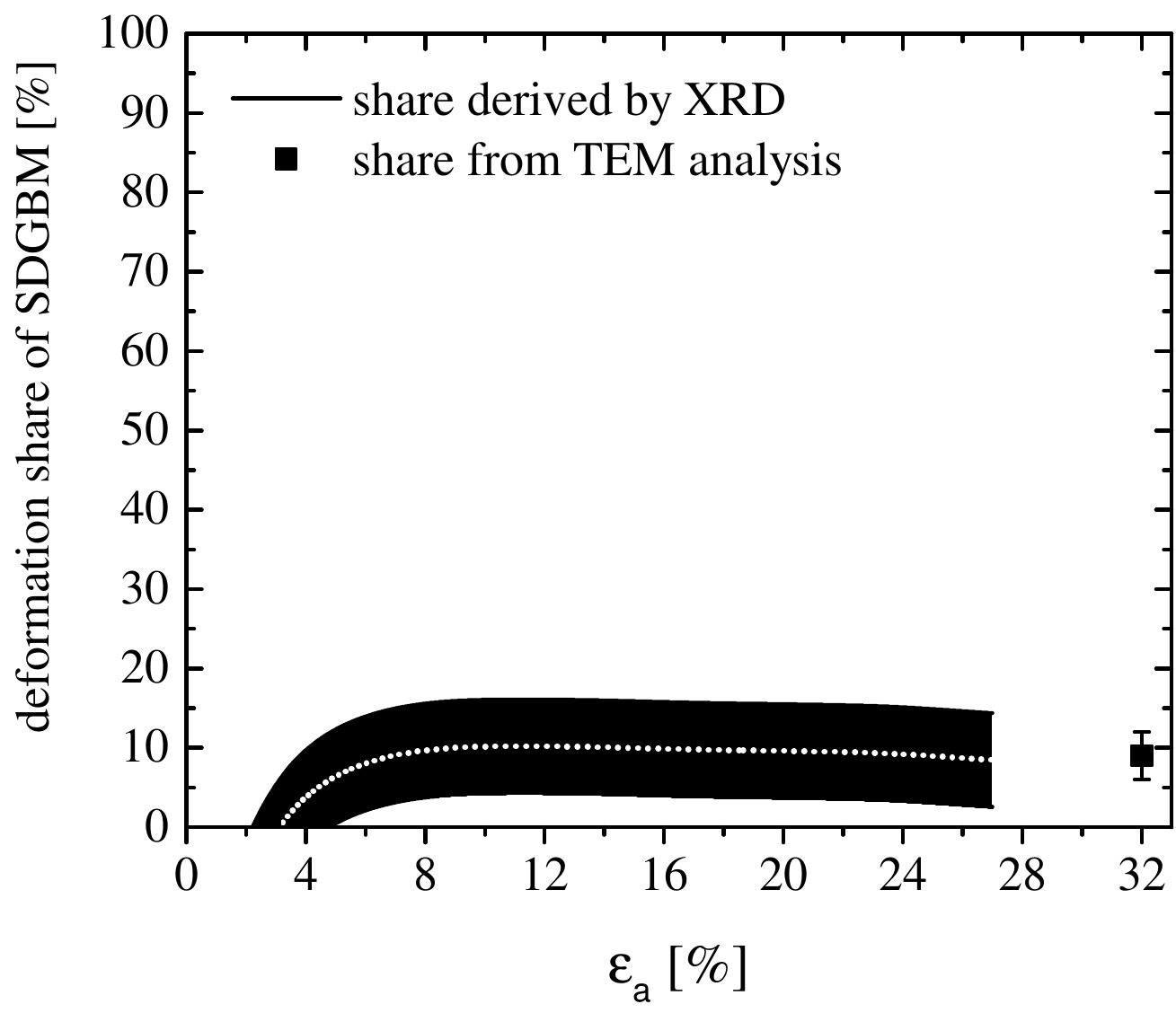}\label{fig:SDGBMShare}}
 \caption[]{(a) Sketch to estimate the shear-strain due to SDGBM. (b) Relative deformation share of SDGBM as a function of applied strain. The single square corresponds to the SDGBM share derived from the TEM grain size histograms shown in Fig. \ref{fig:AspectRatio}.}
 \label{fig:Coupling}
\end{figure*}
%

To estimate the contribution of SDGBM to overall strain, we refer to a simple model displayed in Fig. \ref{fig:CouplingSketch}. We assume growth of the middle grain in response to the applied shear stress $\tau$ which induces a shear-strain $\gamma=\Delta x/3D$. The shear-strain can also be expressed as $\gamma=\beta\Delta D/3D$, where $\Delta D$ denotes the grain growth relative to the initial grain size $D$. Using the $\Dvol$-data in Fig. \ref{fig:GrainSizeEvolution}, an average coupling factor of $ \langle \beta \rangle = 0.3$ \cite{Molodov2011} and the relation $\varepsilon=\gamma/\sqrt{3}$ {\cite{Argon2008}}, it is straightforward to compute the relative contribution $\varepsilon_\mathrm{SDGBM}/\varepsilon_\mathrm{a}$ of SDGBM as a function of applied strain. The resulting evolution of the SDGBM share is shown in Fig. \ref{fig:SDGBMShare} together with the value from the TEM analysis (see Fig. \ref{fig:GSinitial} and \ref{fig:GSdeformed}), which we discuss in the following section. It should be noted that by letting both GBs move in opposite vertical and horizontal directions, as shown in Fig. \ref{fig:CouplingSketch}, we estimate an upper bound for the associated shear-strain.

\FloatBarrier

\subsection{Integral scattering intensity and onset of preferred orientation development}
\begin{figure*}[b!]
\centering
 \hspace*{1.1cm}
     \subfigure[]{\includegraphics[width=0.32\textwidth]{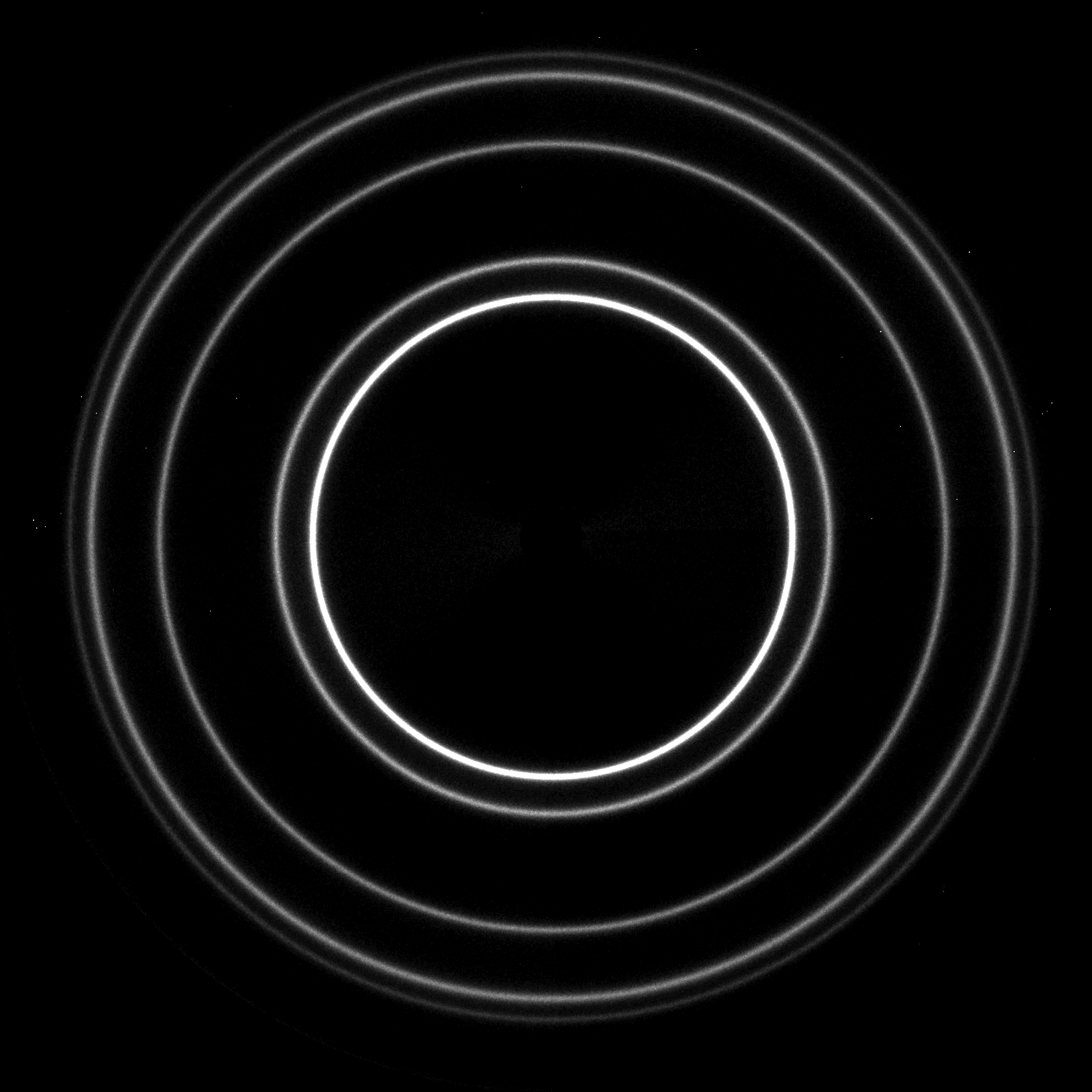}\label{fig:InitialDSRings}}
 \hspace*{1.4cm}
     \subfigure[]{\includegraphics[width=0.45\textwidth]{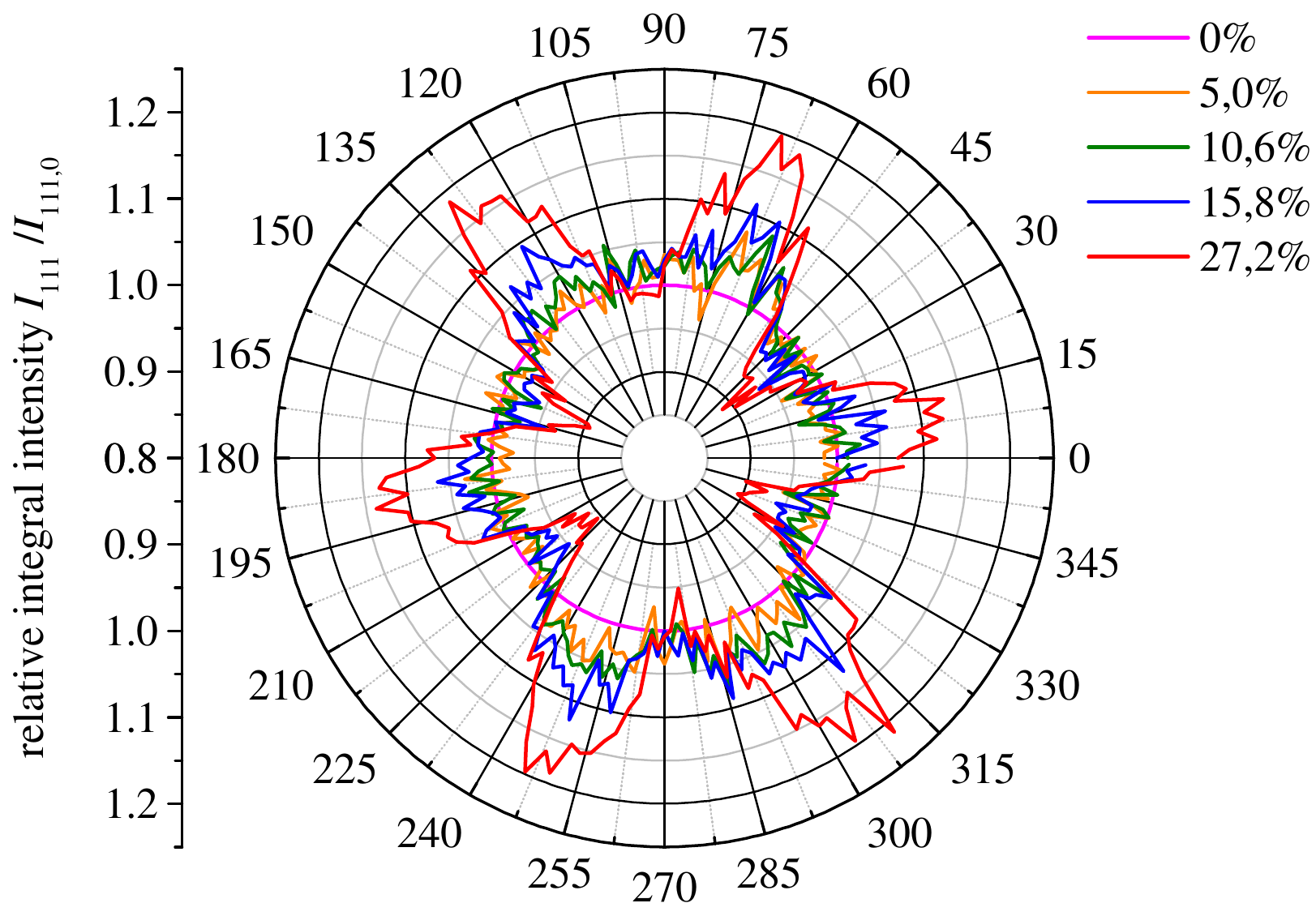}\label{fig:111IntensityPolar}}\\
     \subfigure[]{\includegraphics[width=0.45\textwidth]{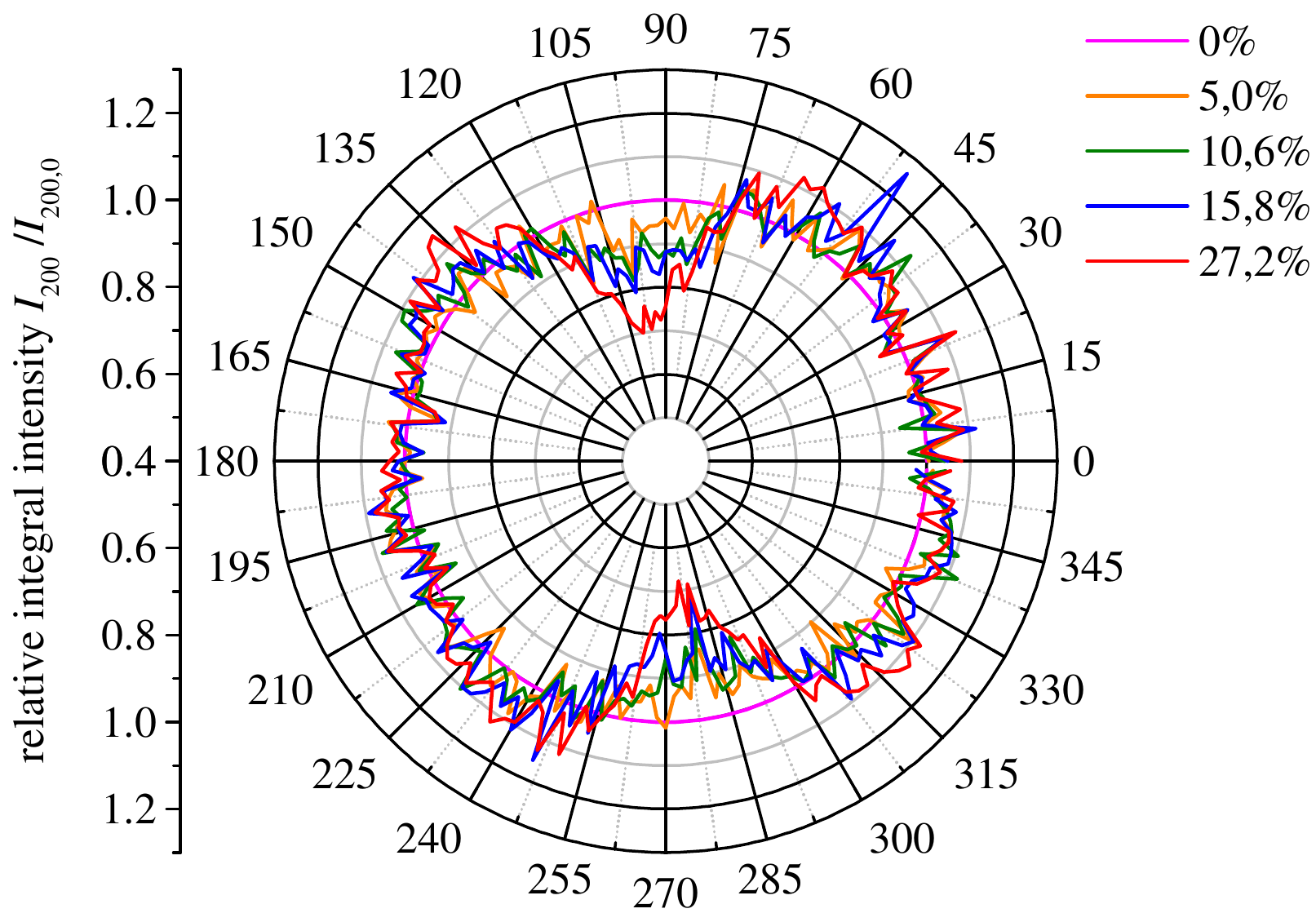}\label{fig:200IntensityPolar}}
 \hspace*{.3cm}
     \subfigure[]{\includegraphics[width=0.45\textwidth]{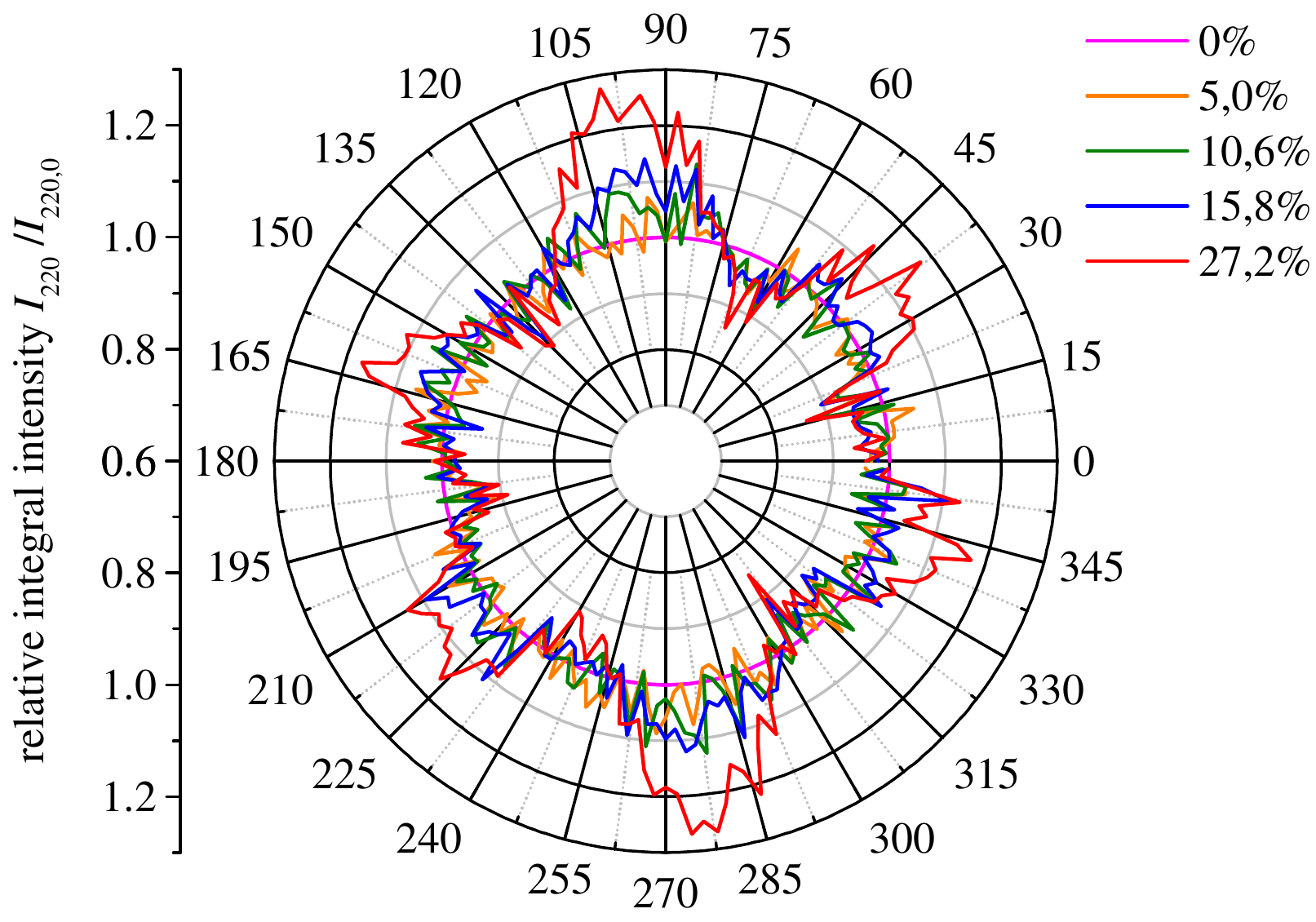}\label{fig:220IntensityPolar}}
 \caption[]{(a) Picture of the detector signal of the initially undeformed NC sample, showing uniformly distributed intensities along the Debye-Scherrer rings, {which are} typical for the scattering of homogeneous and isotropic polycrystals. (b-d) Polar plots of integral scattering intensities at five selected deformation states ((b) \{111\}, (c) \{200\} and (d) \{220\}). Actually, relative intensities $I/I_0$ normalized by the initial scattering intensity $I_0$ of the undeformed SCS are shown to emphasize subtle intensity changes.}
 \label{fig:INTPolar}
\end{figure*}

Before deformation, all \{hkl\}-reflections produced Debye-Scherrer rings of uniform intensity around their circumference (Fig. \ref{fig:InitialDSRings}). This signature is typical for a material made up of randomly oriented grains, i.e. an untextured or statistically isotropic polycrystal. Upon loading, the measured intensities remain fairly constant up to $\varepsilon_\mathrm{a} \approx \unit[2]{\%}$, however, when strain is further increasing a subtle redistribution of scattering intensities is observed as displayed in the polar plots of Fig. \ref{fig:INTPolar}(b-d). These polar plots represent the intensity distributions along the circumference of the related \{111\}-,\{200\}- and \{220\}-pole figures where the shear plane normal (SPN) is oriented along $\unit[135]{^\circ}$ and the shear direction (SD) along $\unit[45]{^\circ}$ (c.f. Fig. \ref{fig:ExpSetup}). A detailed assignment of ideal texture components to the extrema in Fig. \ref{fig:INTPolar} based on \cite{J.-J.Fundenberger2015, Beyerlein2009} is presented in table \ref{tab:texture}.

%
%

%
\begin{table}
\caption{Overview of polar angles of expected intensity extrema along the circumference of pole figures corresponding to \{111\}, \{200\} and \{220\} planes (see fig. \ref{fig:INTPolar}) for the full set of ideal texture components of simple shear in fcc metals.}
\label{tab:texture}
\centering
\begin{tabular}{c c c c c c c }
\hline 
Texture & Shear & Shear & & \multicolumn{3}{c}{Angles of expected extrema of \{hkl\} poles} \\ 
component & plane & direction & \hspace{5mm} & \{111\} & \{200\} & \{220\}  \\
\hline  \\
${A \ / \ \bar{A}}$  & $(11\bar{1}) \ / \ (\bar{1}\bar{1}1)$ & $[1\bar{1}0] \ / \ [\bar{1}10]$ & & $\begin{matrix} 135^{\circ} & 315^{\circ} \end{matrix}$ & -- & $\begin{matrix} 45^{\circ} & 225^{\circ} \end{matrix}$ \\ 
\\
${B \ / \ \bar{B}}$ & $(11\bar{2}) \ / \ (\bar{1}\bar{1}2)$ & $[1\bar{1}0] \ / \ [\bar{1}10]$ & & -- & -- & $\begin{matrix} 45^{\circ} & 225^{\circ} \\  105^{\circ} & 285^{\circ} \\  165^{\circ} & 345^{\circ} \end{matrix} $ \\ 
\\
${C}$ & $(100)$ & $[0\bar{1}1]$ & & $\begin{matrix} 10^{\circ} & 190^{\circ}  \\  80^{\circ} & 260^{\circ} \end{matrix}$ & $\begin{matrix} 135^{\circ} & 315^{\circ} \end{matrix}$ & $\begin{matrix} 45^{\circ} & 225^{\circ} \end{matrix}$ \\ 
\\
${A^{*}_{1}}$ & $(11\bar{1})$ & $[2\bar{1}1]$ & & $\begin{matrix} 25^{\circ} & 205^{\circ} \\ 135^{\circ} & 315^{\circ} \end{matrix}$ & $\begin{matrix} 80^{\circ} & 260^{\circ} \end{matrix}$ & $\begin{matrix} 170^{\circ} & 350^{\circ} \end{matrix}$ \\ 
\\
${A^{*}_{2}}$ & $(1\bar{1}1)$ & $[\bar{2}\bar{1}0]$ & & $\begin{matrix} 65^{\circ} & 245^{\circ} \\ 135^{\circ} & 315^{\circ} \end{matrix}$ & $\begin{matrix} 10^{\circ} & 190^{\circ} \end{matrix}$ & $\begin{matrix} 100^{\circ} & 280^{\circ} \end{matrix}$ \\ 
\\
\hline 
\end{tabular} 
\end{table}
%

Intuitively, one would expect that loading a SCS in the given setup should favor single slip on \{111\}-planes oriented along the SPN and thereby propagating strain toward the SD. It is most surprising that the material does not behave in such a selective manner but rather activates all possible texture-components related to simple shear deformation in a fcc metal, as revealed by table \ref{tab:texture}.

So far we can only speculate that the presence of statistical isotropy of grain orientations before deformation may give rise to activation of all possible texture-components. Selection of primary slip by the highest Schmid-factor therefore seems frustrated. On the other hand, this implies that dislocation activity is rather an accommodation process that goes along with a different but dominating strain carrying deformation mechanism. This view is in line with a recent work of Skrotzki et al. \cite{Skrotzki2013} who studied in great detail texture formation (Euler-space representation) in NC Pd$_{90}$Au$_{10}$ specimens after high pressure torsion (HPT) deformation. For strains up to $\gamma \approx 1$, they observed a random texture undistinguishable from the isotropic microstructure of the undeformed sample state and so argued that GB-mediated processes must carry the deformation without leading to texture formation. In fact, they could not observe the faint changes of relative intensities $I_{\mathrm{hkl}}/ I_{\mathrm{hkl,0}} < 1.2$ discussed above (see Fig. \ref{fig:INTPolar}(b-d)). 

To reconcile their findings with ours, we argue that detection of the emergence of preferred grain orientations at comparatively low strain values with such a high sensitivity is related to the SCS geometry in conjunction with the chosen transmission geometry diffraction set up. In fact, loading of SCS enforces unidirectional plane shear along the gauge section and its effect is in-situ probed by high energy synchrotron radiation penetrating the entire gauge section. Due to the high energy X-ray radiation ($\lambda = \SI{0.178}{\angstrom}$), the scattering vector is oriented nearly perpendicular to the incoming beam direction so probing all possible directions ($0^{\circ} \leq \phi < 360^{\circ}$) in the plane of the gauge section. In contrast to that, Skrotzki et al. utilized position-sensitive X-ray microdiffraction equipped with an Eulerian cradle and a two-dimensional detector on a lab diffractometer operated in reflection geometry ($\mathrm{Cu K}_{\alpha}$-radiation, $\lambda = \SI{1.542}{\angstrom}$). Their samples were also prepared by IGC, employing the same processing protocol as discussed in section 2. However, they probe the near surface region of the HPT-deformed sample since the penetration depth of $\mathrm{Cu K}_{\alpha}$-radiation is of the order of $\unit[1]{\mu m}$ in Pd-Au alloys. Further contributions to the observed discrepancy may relate to the fact that in-situ deformation and post-deformation studies unveil diverse behavior. Likewise, the two different diffraction set-ups may operate with different sensitivity when recording intensity changes along the circumference of a pole figure. 

To complete the discussion, we note that Skrotzki et al. \cite{Skrotzki2013} also investigated texture formation in the large strain regime ($1< \gamma <17$), to find the evolution of a ``brass-type'' texture which mainly consists of $B/\bar{B}$ simple shear components. Texture modeling revealed that best agreement with experiment is obtained by referring to the Taylor model in conjunction with the $\{111\}\langle {112} \rangle$ slip system family which is by far dominating strain propagation. Moreover, they argue that slip should relate to nucleation and emission of $1/6 \langle {112} \rangle$ partial dislocations from GB sources. Generally, slip of such partials should generate stacking faults in the grains. By peak profile analysis they found however that stacking faults and twins have a rather low density (present in 2-3\% of the grains). As a result, they concluded that intrinsic stacking faults become eventually eliminated by a time-shifted trailing partial emitted from the same GB source.

{Despite the rather complex evolution of texture components in the low strain regime ($\gamma < 0.3$), we notwithstanding concentrate now on estimating an upper bound for dislocation mediated plasticity based on change of grain shape associated with texture formation. To correlate applied strain with change of aspect ratio, we invoke a simple model conception that approaches grain shape by an ellipse and allows for shearing it into a given direction and so generating shape anisometry.}

\begin{figure}[t!]
     \subfigure[]{\includegraphics[width=0.45\textwidth]{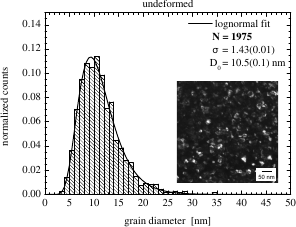}\label{fig:GSinitial}}
 \hspace*{.3cm}
     \subfigure[]{\includegraphics[width=0.45\textwidth]{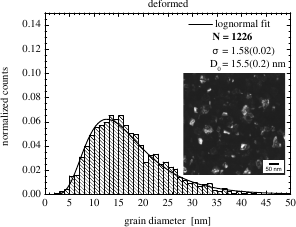}\label{fig:GSdeformed}}\\
     \subfigure[]{\includegraphics[width=0.45\textwidth]{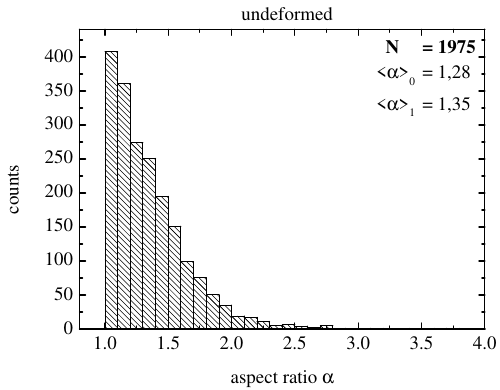}\label{fig:ARinitial}}
 \hspace*{.3cm}
     \subfigure[]{\includegraphics[width=0.45\textwidth]{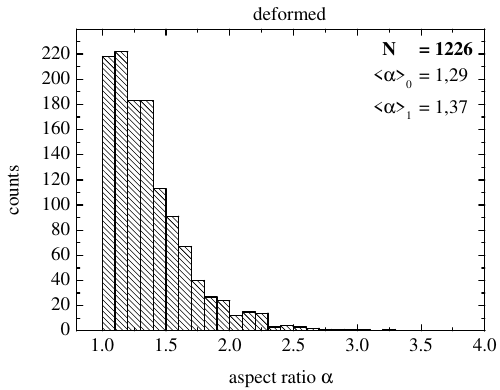}\label{fig:ARdeformed}}
 \caption[]{Grain size and aspect ratio distributions of a Pd$_{90}$Au$_{10}$-SCS measured by TEM. {The diagrams (a, c) shows the data of the elastically deformed bulk of the specimen and (b, d) of the gauge section deformed to $\unit[32]{\%}$ applied strain. In (a, b) the arithmetic mean of the longest and shortest chord axis was used to generate grain diameter histograms which were fitted by a log-normal distribution to find the median $D_0$ and distribution width $\sigma$.} In case of the aspect ratio, the median $\langle\alpha\rangle_0$ and the arithmetic mean $\langle\alpha\rangle_1$ were calculated directly from the histogram. In all cases, $N$ denotes the number of counted grains. The insets in (a) and (b) show typical TEM dark field micrographs.}
 \label{fig:AspectRatio}
\end{figure}

We define aspect ratio $\alpha$ of grains as the ratio of {chord length} measured along the short and long axis of the cross-section of individual grains flashed up in TEM dark-field micrographs. These were obtained from thin lamellae taken from the only elastically deformed bulk of the specimen as well as from the gauge section having experienced $\approx\unit[32]{\%}$ applied strain (for details see section 2). The so recorded data are displayed in Fig. \ref{fig:AspectRatio}. The aspect ratio histogram of the undeformed material is shown in Fig. \ref{fig:ARinitial}, and the histogram of the plastically deformed gauge section is presented in Fig. \ref{fig:ARdeformed}. {Related to the faint changes of relative intensities $I_{\mathrm{hkl}}/ I_{\mathrm{hkl,0}} < 1.2$, which are indicative of the onset of preferred orientation formation, we observe only a small change of the arithmetic mean value of the aspect ratio distributions of the undeformed ($\langle\alpha\rangle_1\approx 1.353 \pm 0.003$) and deformed ($\langle\alpha\rangle_1\approx 1.366 \pm 0.004$) material. The error $\Delta \langle\alpha\rangle_1$ of the respective arithmetic mean values were computed based on the precision of measuring the longest and shortest chord length of individual grains ($\pm \unit[0.5]{nm}$, see section \ref{Material-and-Methods} Materials and methods). The corresponding maximal standard deviation $\Delta\alpha$ follows from the standard formula for linear error propagation. As a result, $\Delta \langle\alpha\rangle_1 = \Delta\alpha / \sqrt{N}$, where $N$ is the number of measured grains indicated in Fig. \ref{fig:AspectRatio}.}

%
%

Shear-strain through dislocation glide will be assessed with the idea that dislocations incrementally contribute to the change of aspect ratio (grain elongation along the glide direction and shrinkage in the respective perpendicular direction). For example, when dislocations (or pairs of partials) traverse a crystallite of equivalent diameter $D$, it is sheared by a shear increment $\delta s$. This causes a shear-strain $\gamma\left( \delta s,D \right)\approx \delta s/D$ respectively the related plastic strain $\varepsilon_\mathrm{p,dis} \left( \delta s,D\right)=\gamma/\sqrt{3}$ \cite{Argon2008}. Estimating the associated change of aspect ratio $\alpha$ of grains relies upon their geometry. Since we apply plane shear, let us assume for the sake of simplicity that the initially undeformed grains can be represented by their two dimensional cross section. We further simplify matters by approximating the cross section by an ellipses with major semiaxis $A$ and minor semiaxis $B$; this geometry is commonly used by automated crystal orientation mapping by means of TEM (ACOM-TEM) \cite{Kobler2013,Kobler2015} to represent grain-shape information. Complying with the definition of grain size $D$, the ellipse has identical area as a circular grain with diameter $D$, i.e. $ \pi D^2/4 = \pi A B$ and the aspect ratio $\alpha = A/B$. 

Using cartesian coordinates, an ellipse, with its center at the origin $(x = y = 0)$, is represented by the following parametric equation
\begin{align}
	\vec{E}(t) = \begin{pmatrix}
		x(t) \\[0.5em]
		y(t) 
	\end{pmatrix}\ &= \begin{pmatrix}
		A \cos\left( t \right) \cos\left(\varphi\right) - B \sin \left( t \right) \sin \left( \varphi \right)  \\[0.5em]
		A \cos\left( t \right) \sin\left(\varphi\right) - B \sin \left( t \right) \cos \left( \varphi \right)
	\end{pmatrix} + \begin{pmatrix}
		0	\\[0.5em]
		\frac{\delta s}{A} \left( A \cos\left( t \right) \cos\left(\varphi\right) - B \sin \left( t \right) \sin \left( \varphi \right) \right)
	\end{pmatrix} \label{equ:ellipse1}
\end{align}
with the parameter $t \in \left[ 0,2\pi \right)$ and $\varphi$ which is the angle between the x-axis and the major semiaxis. The first term represents the unsheared ellipse, and the second term accounts for shearing of the ellipse in y-direction by $\delta s/A$, where $s$ is a scalar displacement variable. The quantity $\arrowvert \vec{E}(t) \arrowvert$ denotes the distance between a point on the contour of the ellipse, specified by $t$, and the origin of the coordinate system. Therefore, the maximum and minimum of $\arrowvert \vec{E}(t) \arrowvert$ straightforwardly provide the length of the major and minor semiaxes of the sheared ellipse, which then allows us to compute the shear-dependent aspect ratio $\alpha$. 

Instead of using multiple ellipses to approximate individual grains, we approximate the aggregate of grains by a single representative ellipse. It is essentially generated by the mean aspect ratio $\langle \alpha \rangle$ and the mean grain size $\langle D \rangle$. Consequently, the scalar shear displacement acting on the ellipse also epitomizes a mean displacement $\langle \Delta s \rangle$. The central task is now to reproduce the experimentally extracted change of mean aspect ratio $\langle \alpha \rangle_{1}$ by applying a suitable amount of shear displacement to the representative ellipse. The required shear displacement $\langle \Delta s \rangle$ is exclusively attributed to dislocation activity and hence represents an upper bound of plastic deformation due to dislocation glide. 

\begin{figure*}[t]
     \subfigure[]{\includegraphics[width=0.425\textwidth]{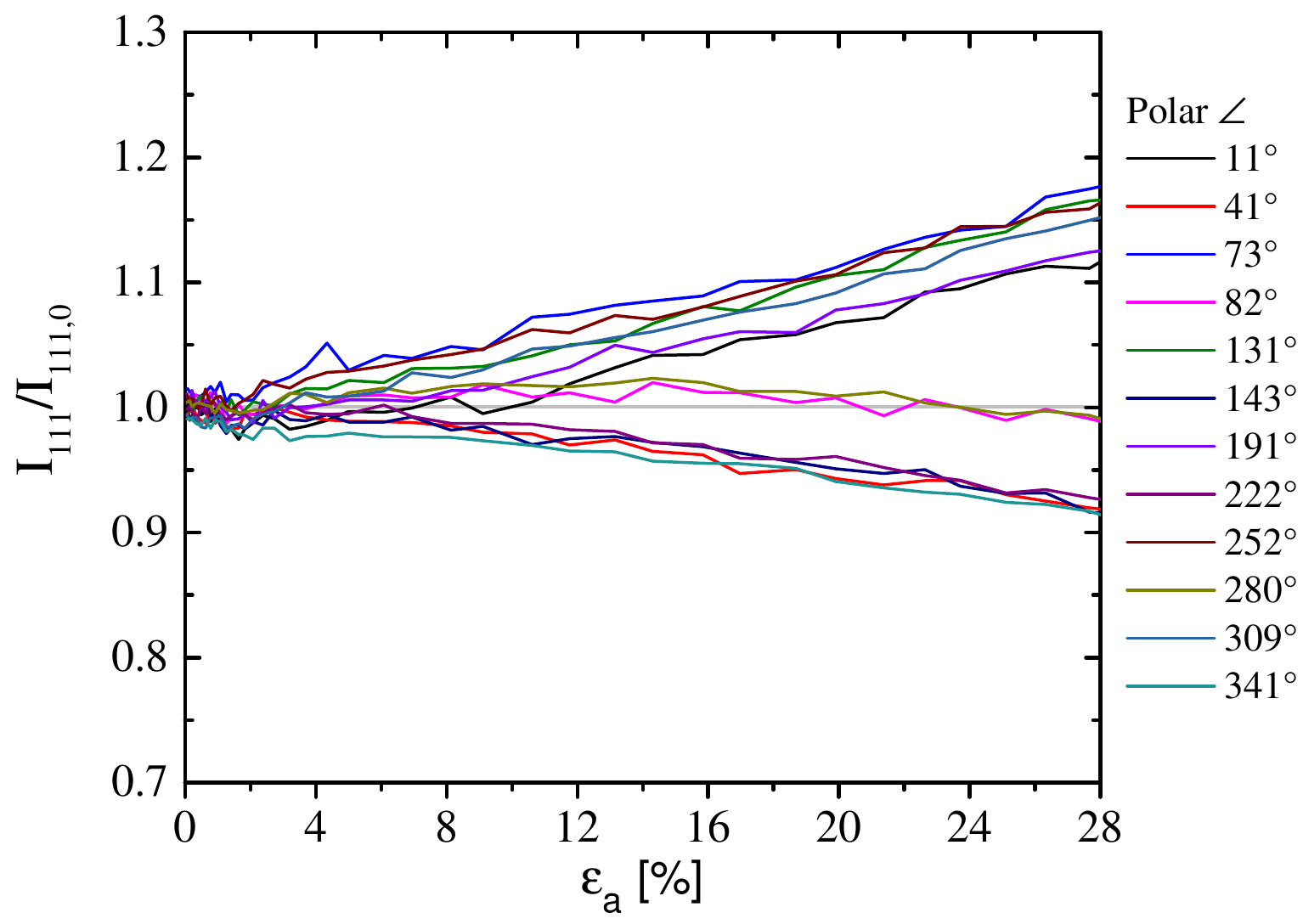}\label{fig:Rel_Int_111}}
 \hspace*{.3cm}
     \subfigure[]{\includegraphics[width=0.425\textwidth]{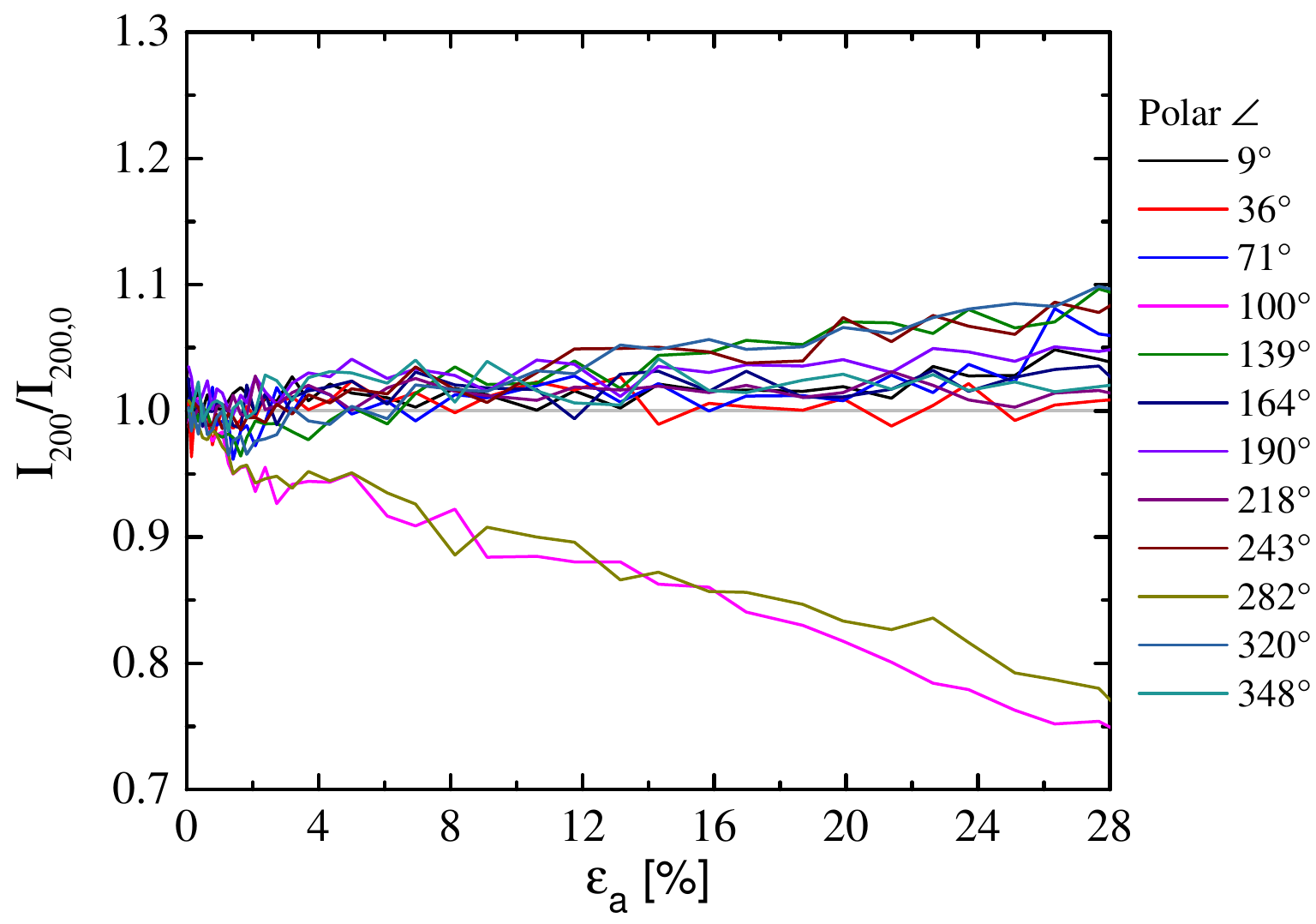}\label{fig:Rel_Int_200}}\\
     \subfigure[]{\includegraphics[width=0.425\textwidth]{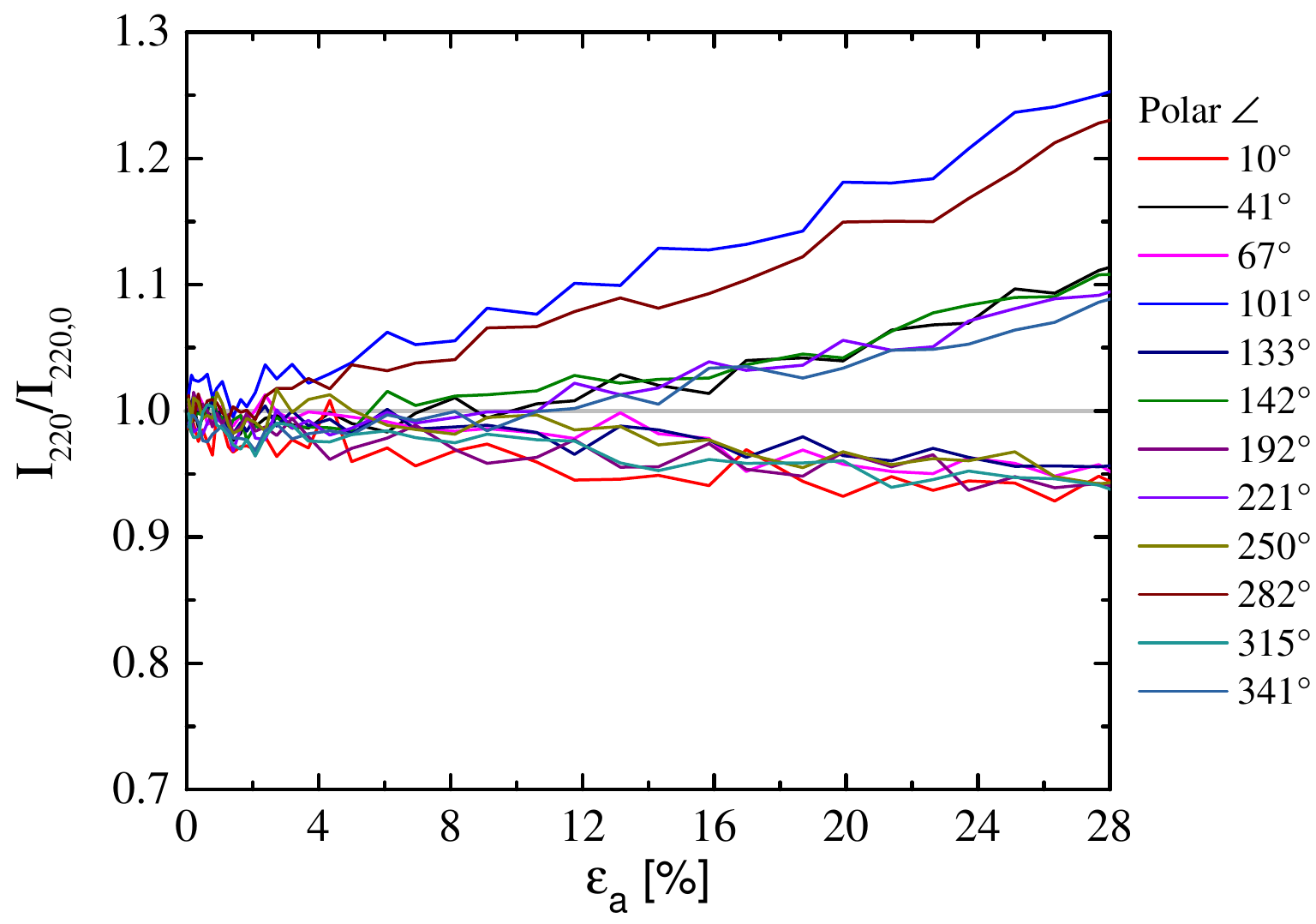}\label{fig:Rel_Int_220}}
 \hspace*{.36cm}
     \subfigure[]{\includegraphics[width=0.358\textwidth]{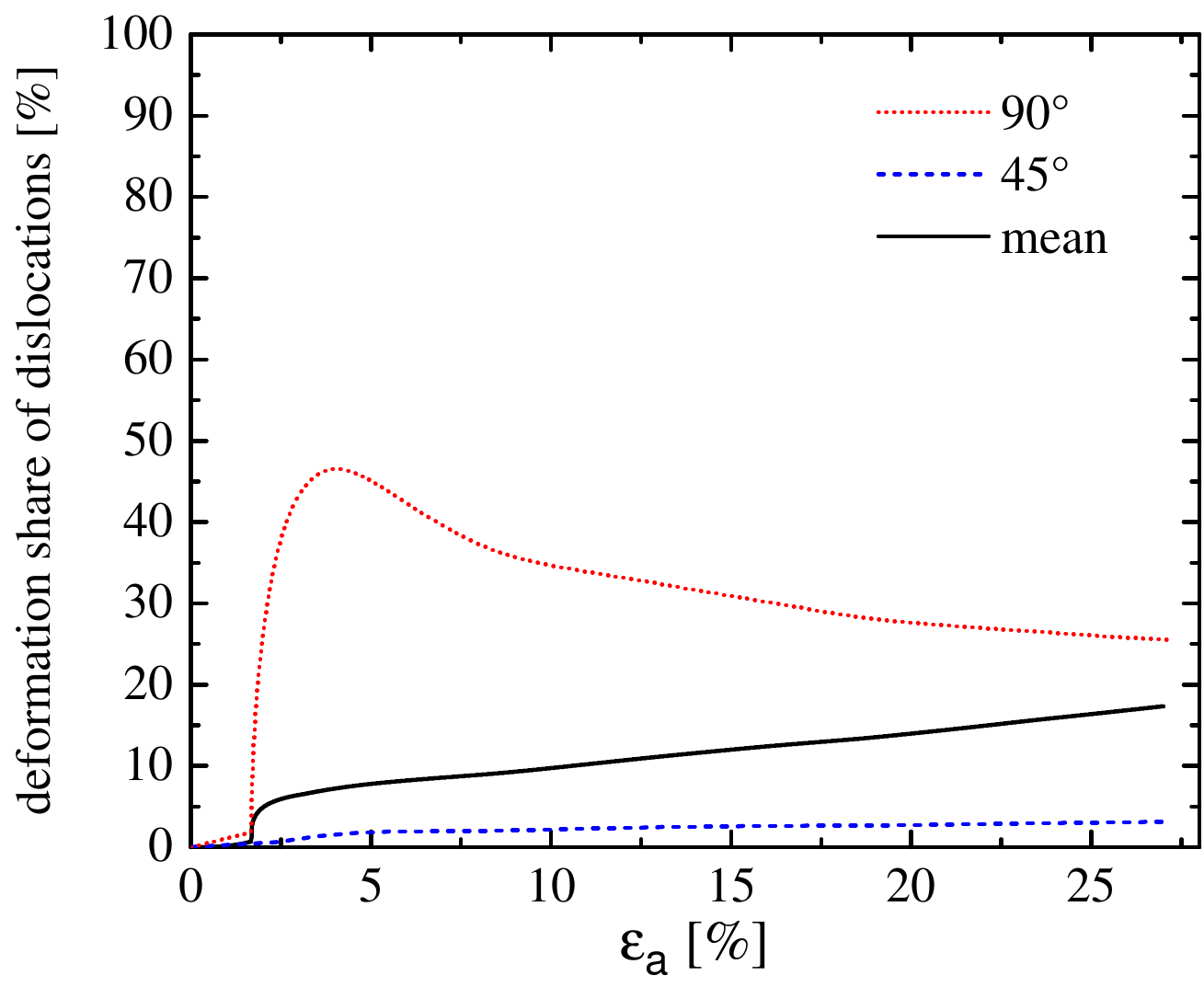}\label{fig:DislocationEllipses}}
 \caption[]{Evolution of the normalised scattering intensity at the extrema of the polar plots at distinct polar angles for (a) \{111\}-, (b) \{200\}- and (c) \{220\}-planes. (d) Dislocation share for different starting angles $\varphi$ of sheared ellipses. The mean value is averaged over $0^{\circ} \leq \varphi \leq 90^{\circ}$ and corresponds to a random distribution of starting angles.}
	\label{fig:DislocationShareFromINTInc}
\end{figure*}

To correlate change of aspect ratio and strain supplied by dislocation activity, we need to interpolate between the known aspect ratios of the undeformed and deformed state. As a working hypothesis, it is assumed that the aspect ratio increases linearly with applied strain. This premise is supported by the following evidence: the strain-induced evolution of relative integral intensities of the \{111\}-, \{200\}- and \{220\}-poles manifest a basically linear increase/decrease that starts at $\varepsilon_\mathrm{a} \approx \unit[2]{\%}$ and is independent of the considered texture components, as displayed in Fig. \ref{fig:DislocationShareFromINTInc} for relative intensity extrema of all texture components. Having thus substantiated the linear interpolation scheme, it is now straightforward to correlate applied strain $\varepsilon_{\mathrm{a}}$ with mean aspect ratio $\langle\alpha\rangle_1$ and associated mean shear displacement $\langle \Delta s \rangle$.

%
%

There is a second aspect that may cause concern. The grain size increases simultaneously during deformation, although \textit{not }developing grain shape anisometry (see Fig. \ref{fig:GrainSizeEvolutionPolar} and the discussion in subsection \ref{Integral-peak-width}). However, grain growth definitely affects the length of the major and minor semiaxes and therfore impacts the strain $\varepsilon_\mathrm{p,dis}\left(s,D \right)$ associated with dislocation glide. To account for this circumstance, we chose an iterative approach where an ellipse is stepwise sheared by a displacement increment $\langle \delta s \rangle$. After each shear increment, the resulting aspect ratio $\langle \alpha \rangle$ is computed and correlated with the total applied strain $\varepsilon_\mathrm{a}$ and grain size $\langle D \rangle$. With this set of parameters a new ellipse is created which then experiences the next shear increment. This procedure is repeated as often as necessary to cover the full range of applied strain.

The fact that we treat $\varphi$ and $s$ as continuous variables accounts for the operation of the full set of texture components in a randomly oriented aggregate of grains. Consequently, we do not rely upon dislocation activity in a primary slip system with well-defined orientation relationships of glide plane and direction to SPN and SD. In particular, we find that the initial orientation of the unsheared ellipse relative to the direction of shear has a tremendous impact on the mean displacement $\langle \Delta s \rangle$ needed to reach a given change in aspect ratio $\langle\alpha\rangle_1$ and, therefore, amount of strain caused by dislocation glide or likewise its relative contribution, $\varepsilon_\mathrm{p,dis}/\varepsilon_{\mathrm{a}}$, to total applied strain. The limiting cases ($\varphi = 90^{\circ}$ and $\varphi = 45^{\circ}$) are shown in Fig. \ref{fig:DislocationEllipses}. To estimate a pragmatical bound of dislocation glide to total applied strain, we average over a random distribution of grain shape orientations $\varphi \in \left[0^{\circ},90^{\circ}\right]$, which is sampled in $\unit[0.5]{^{\circ}}$ steps. The resulting mean share of dislocation glide is shown in Fig. \ref{fig:DislocationEllipses} and reveals an abrupt increase at $\varepsilon_\mathrm{a} \approx \unit[2]{\%}$ followed by a sluggish monotonic increase. Below $\varepsilon_\mathrm{a}=\unit[11]{\%}$, the dislocation share contributes less than the elastic lattice strain to total deformation and even at the highest strain in the experiment, $\varepsilon_\mathrm{a}=\unit[27]{\%}$, its contribution to overall strain does not exceed $\unit[20]{\%}$. 

Interestingly, \textit{Bachurin} et al. \cite{Bachurin2010} carried out atomistic simulations of uniaxial tensile and compressive loading of NC Pd with a mean grain size of $\unit[10]{nm}$. They came to the conclusion that the contribution of extended partial dislocations emitted from the GBs as well as full dislocations and twinning (at later stages of deformation) to the total strain is insignificant. Summing up over all sorts of dislocation contributions including dislocation embryos, they found at a total strain of $\approx \unit[10]{\%}$ a relative share of dislocation-based deformation of about $\unit[10]{\%}$.

As a summary of this section, we conclude that in NC metals with a mean structural correlation length of $\unit[10]{nm}$ dislocations rather serve as carrier of accommodation processes instead of acting as generic flow defect which propagates strain in a dominant manner.

\section{Conclusions}
%
\begin{figure*}[t]
     \subfigure[]{\includegraphics[width=0.46\textwidth]{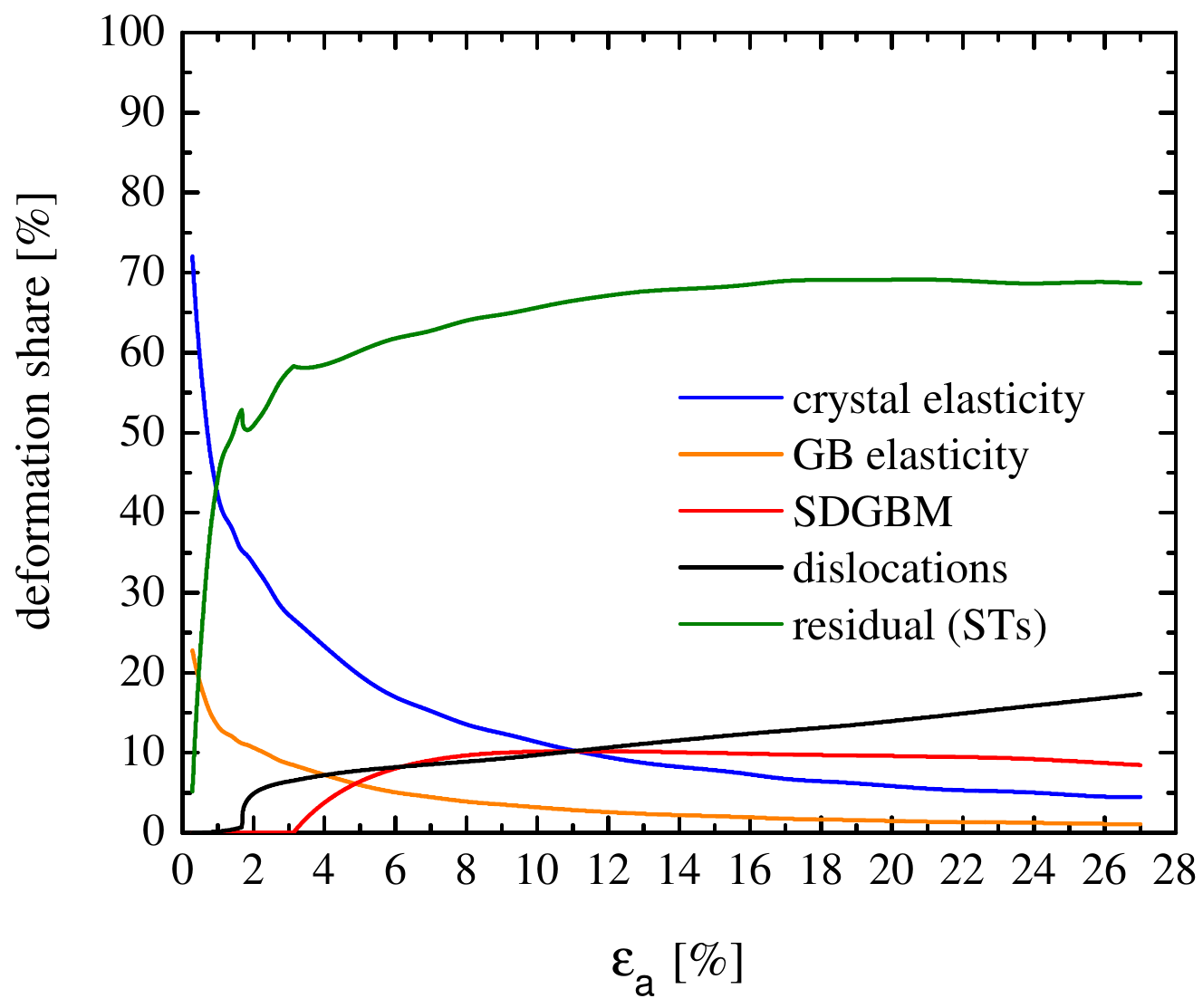}\label{fig:Contributions}}
 \hspace*{.3cm}
     \subfigure[]{\includegraphics[width=0.45\textwidth]{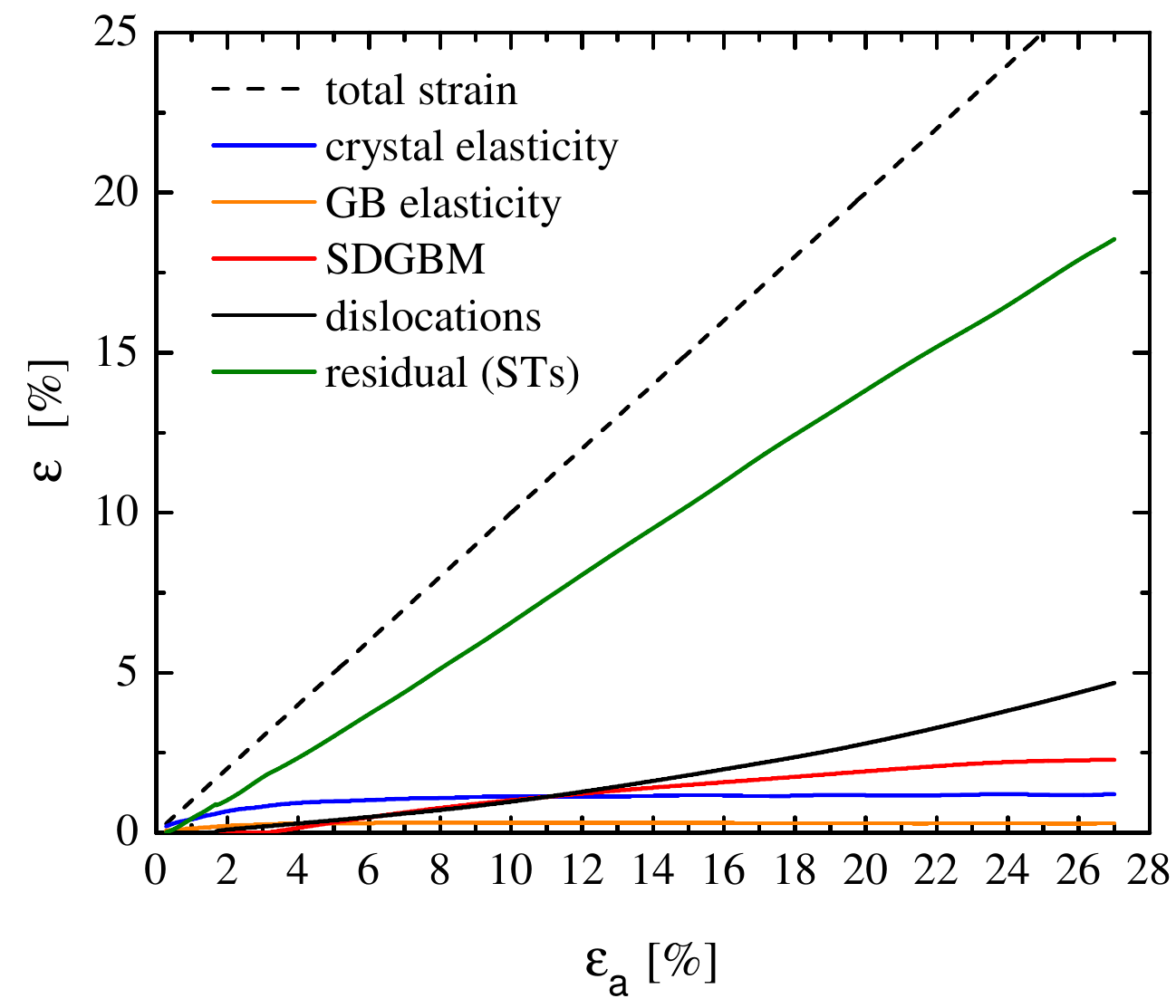}\label{fig:Contributions2}}
 \caption[]{(a) Contributions of different deformation mechanisms relative to the applied strain $\varepsilon_\mathrm{a}$. The green line shows the evolution of the residual share, which we assign to ST-induced accumulation of shear-strain at/along GBs. (b) Strain contributions of different deformation mechanisms to the total applied strain (dashed line). This representation clarifies, that all deformation mechanisms continuously contribute to the overall deformation, even though their relative share may decrease (c.f. elasticity or SDGBM).}
 \label{fig:2contributions}
\end{figure*}
%

In the previous sections we identified four different strain carrying mechanisms as well as their relative contributions to overall deformation, which are altogether displayed in Fig. \ref{fig:Contributions}. In addition to the elastic, dislocation based, and SDGBM strain shares, we also show the difference between the sum of these shares and \unit[100]{\%} strain contribution. Strikingly, this residual contribution, steeply rising in the early stage of deformation, approaches a strain share of more than \unit[60]{\%} implying that about two thirds of total deformation are yet unaccounted for. Since X-ray diffraction is particularly sensitive to microstructural changes associated with deviations from perfect intracrystalline order but is non-sensitive to intercrystalline interference, it seems self-evident that the unaccounted strain contribution must have been propagated by GB-mediated deformation modes. The core structure of GBs and possible deformation-induced configurational changes of GB cores give rise to diffuse scattering but its analyis is highly involved and therefore beyond the scope of this study.

In a recent work \cite{Grewer2014}, we have been led to the same conclusion, however, based on reverse reasoning. By deriving expressions for the \textit{effective} activation volume and utilizing experimental or theoretical values for the specific activation volumes of the involved deformation mechanisms, we estimated the relative contributions of deformation modes to arrive at a share of GB-mediated deformation through ST-activity of $\approx \unit[70]{\%}$. We note that the GB-mediated deformation share obtained from in-situ deformation and diffraction has lower bound character in strict sense, insofar both ways of reasoning agree. From the determination of the effective activation energy and its stress dependence \cite{Grewer2014}, we concluded that shear transformations (STs), the generic flow event in metallic glasses, must operate in/at GBs. We notice that in contrast to metallic glasses, where emerging STs will eventually self-organize to form macroscopic shear bands, the topology and connectivity of GBs effectively impede the emergence of shear bands and thus prevent the material from stick-slip behavior or catastrophic failure (Fig. \ref{fig:StressStrain}). Based on this independent evidence, it seems justified now to maintain that STs, emerging in the core region of GBs, account for about two thirds of total strain in NC Pd$_{90}$Au$_{10}$ at the low end of the nanoscale ($D \leq \unit[10]{nm}$). Likewise, in a recent study \cite{Leibner2015} we demonstrated that plastic yielding in NC Pd-Au alloys mimics universal behavior of metallic glasses, entailing a linear correlation between shear strength and shear modulus. As a consequence, one observes solid solution softening of NC Pd-Au alloys (mean grain size $\leq \unit[10]{nm}$) with increasing Au-concentration in contrast to hardening seen in the coarse grained counterparts. We, therefore, suggested that STs rather than dislocations are the dominant carriers of strain in the fully miscible NC Pd$_{x}$Au$_{1-x}$, $100 \geq x > 50$, alloy system. In fact, predictions of the change of strength of the NC Pd-Au alloy system, which have been derived from available dislocation-based models and theories, could be disproved without exception. 

{In the pertinent literature dealing with nanoplasticity, GB-mediated deformation frequently becomes associated with GB sliding. This term has been originally introduced to denote the rigid body translation of abutting crystallites along a shared interface that produces offsets in marker lines at the GBs. The two classical modes of GB sliding comprise Rachinger sliding, which must be accommodated by intragranular dislocation glide and climb, and Lifshitz sliding that is based on stress-directed diffusion of vacancies and is self-accommodating. These sliding modes occur under creep conditions at elevated temperatures and are characterized by strain-rate sensitivities $> 0.3$ and activation volumes $< \unit[0.1]{b^3}$, respectively. By comparison with activation parameters from our previous work \cite{Grewer2014}, GB sliding can be ruled out being active in NC PdAu at the given testing conditions. In line with this reasoning, Wu et al. \cite{Wu2013} deduced from computer simulation on NC Ni that the occurence of rigid body translation of grains is related to the system size. When it is not large enough, the proximity of free surfaces reduces the constraint exerted by surrounding grains so favoring easy sliding; the opposite is true for increasing system size. Thus, they argued that in realistic NC microstructures GB sliding is kinematically constrained by triple junction lines and quadrupole points. Nevertheless, STs evolving in the core region of GBs are not hampered by these constraints, in fact, their occurence becomes even facilitated since the shear modulus of GBs in NC metals is reduced by $\unit[30]{\%} - \unit[50]{\%}$ \cite{Grewer2011,Leibner2015} compared to the respective bulk values. On the other hand, dislocation shear, which is nucleation controlled at the low end of the nanoscale \cite{Bachurin2010}, therefore, should exert comparatively larger resistance to shear deformation. For that reason, we prefer to assign the irreversible shear-strain accumulated along/at GBs to STs and omit the term GB sliding to avoid confusion with nomenclature in what follows. We finally notice that the activity of STs in the core region of GBs genuinely allows for translational and rotational degrees of freedom of grains relative to each other and so is not in conflict with the emergence of displacement discontinuities across/along GBs to maintain compatibility.}

%
\begin{figure}[t]
	\centering
		\includegraphics[width=0.45\textwidth]{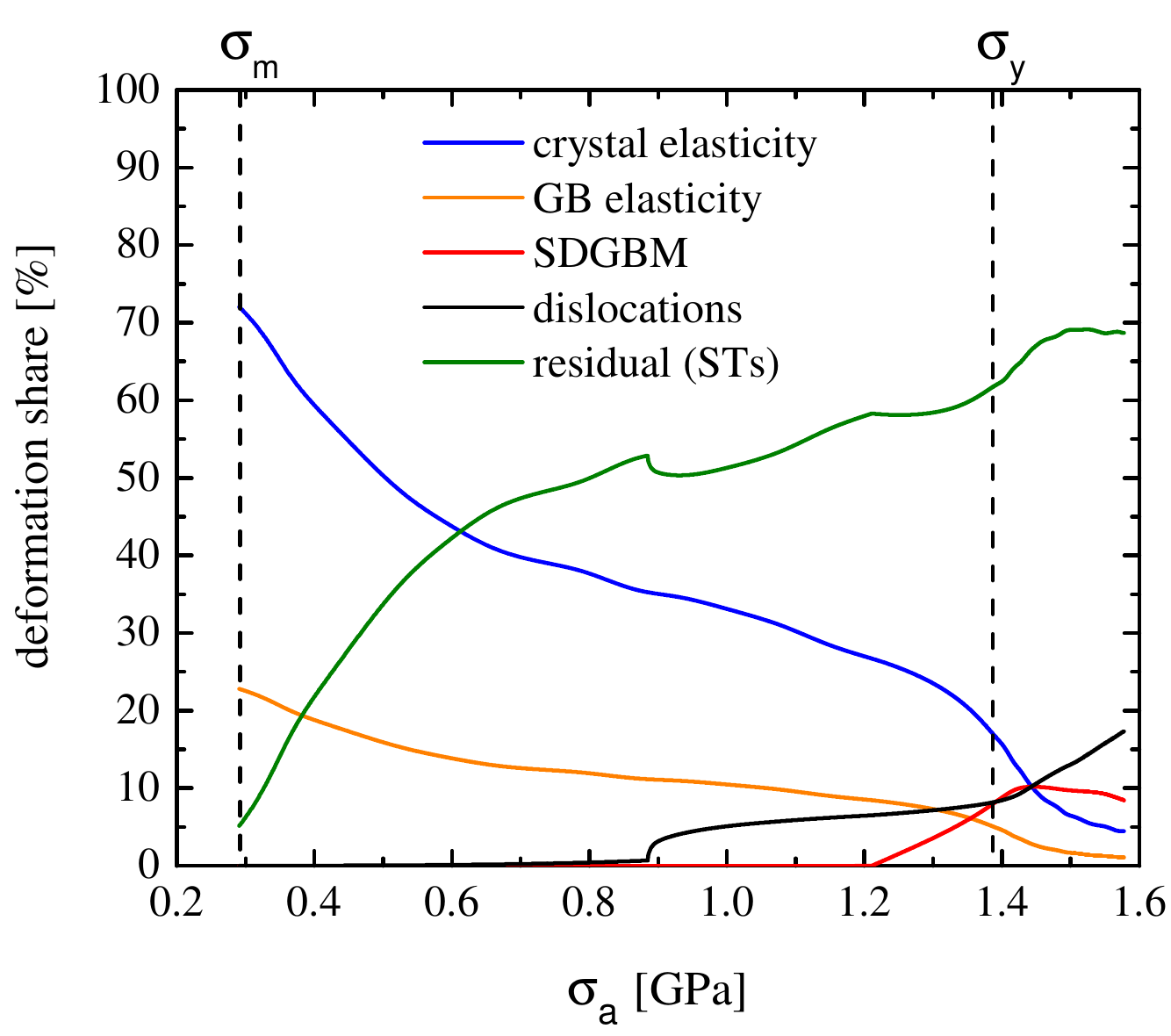}
	\caption{Contributions of different deformation mechanisms relative to the applied stress $\sigma_\mathrm{a}$. Just like in Fig. \ref{fig:Contributions}, the green line shows the evolution of the residual share, which we assign to ST-mediated deformation.}
	\label{fig:ContributionsStress}
\end{figure}
%
{In Fig. \ref{fig:ContributionsStress}, we display the shares of the different mechanisms as a function of applied stress to unravel their simultaneous and/or sequential onset and persistence. For convenience, we also designate the characteristic stress values extracted from yield criteria applied in the stress-strain domain \cite{Ames2012}: $\sigma_\mathrm{m}$ marks the deviation from linear stress-strain behavior, the onset of so-called microplasticity, and $\sigma_\mathrm{y}$ denotes the yield stress, agreeing with the stress value at which the applied strain rate approaches the material´s intrinsic plastic strain rate. We omit the data below $\sigma_\mathrm{m}$ since sample mounting problems generated uncertainties in this regime of the load displacement curves (see section 2). To nevertheless shed light on this deformation stage, computer simulations unraveld that loading a specimen along a perfectly linear stress-strain trajectory ($\sigma < \sigma_\mathrm{m}$) does not necessarily imply presence of Hookean elasticity. Bachurin et al. \cite{Bachurin2012} demonstrated that unloading a sample in the \textit{apparent} elastic regime already results in non recoverable inelastic strain which amounts to $0.1 - 0.15 \%$ at an applied strain of 1\%. This observation is consistent with our finding that the share of the residual contribution in Fig. \ref{fig:ContributionsStress} assumes a small but finite value at $\sigma_\mathrm{m}$.} 

In the stress regime $\sigma_\mathrm{m} < \sigma_\mathrm{a} < \sigma_\mathrm{y}$ the deformation behavior is dominated by Hookean elasticity of the grains and GBs as well as shear shuffling (STs) taking place in the core region of GBs. We observe a rather abrupt onset of dislocation activity at around $\unit[0.9]{GPa}$ followed by a rather weak increase up to $\sigma_\mathrm{y}$. The onset of SDGBM starts out at around $\unit[1.2]{GPa}$ to approach a maximum slightly above $\sigma_\mathrm{y}$. Interestingly, the yield stress determined by prescribing that the imposed strain rate compares with the intrinsic plastic strain rate obviously ensures that the full set of deformation modes is active at $\sigma_\mathrm{y}$. Ignoring for the sake of argument the contributions of SDGBM and dislocations which become effective above $\approx \unit[1.0]{GPa}$, our finding has two profound implications. First, GB-mediated deformation by STs is capable of propagating plastic strain without assistance/interplay of simultaneously acting and irreversible deformation carrying accommodation mechanisms. This behavior is in particular relevant in the stress regime below $\unit[1.0]{GPa}$. When analyzing the activation energy of shear shuffling, Grewer et al. {\cite{Grewer2014}} had to explicitly assume that the activity of STs is \textit{not} coupled to accommodation mechanisms with activation energies distinctively deviating from activation of STs. The identified sole activity of STs acting as carriers of inelastic strain and their evolution to become the dominant strain carrier at large strain thus a posteriori justify the presumption made in \cite{Grewer2014}. Secondly, the conventional concept of work or strain hardening that essentially refers to intracrystalline dislocation interactions seems not being applicable here. This notion also coincides with the analysis of the strain-rate dependence of stress in \cite{Grewer2014}, where it could be verified that the material response in the stress regime $\sigma_\mathrm{m} < \sigma_\mathrm{a} < \sigma_\mathrm{y}$ agrees with nonlinear viscous behavior. 

When $\sigma_\mathrm{a} > \sigma_\mathrm{y}$ the share of SDGBM runs through a maximum to then decrease, whereas, dislocation activity gains progressively more influence to develop the largest rate of increase when the largest stress values are reached. Hookean lattice elasticity tends to assume the minimum share ($\approx 5\%$) here and STs carry a strain share of about 70\%. It seems conceivable that SDGBM and dislocation glide may interact in a synergistic manner e.g. through enhanced grain growth and concomitant increasing intracrystalline dislocation activity at local stress concentrations thereby avoiding early brittle fracture. Overall, the diverse evolution of the respective deformation shares appears as being just a consequence of a complex hierarchical order of onset/nucleation and upholding stresses of different deformation mechanisms, the presence and degree of activity of which strongly depends on grain size. For example, in a recent study of NC Ni with a grain size of \unit[30]{nm}, a strain share of about \unit[40]{\%} could be assigned to dislocation activity \cite{Lohmiller2014}. We finally conjecture that the state of the GB core structure, whether in relaxed (low-excess-energy) or as-prepared (high-excess-energy) configuration, should also have a decisive effect on the share and onset stresses of the above discussed deformation modes. Furthermore, the stacking fault energy of the Pd-Au alloys decreases with rising Au concentration \cite{Schafer2011,Jin2011} and, therefore, partial dislocation activity should increase towards Au-rich alloys. Consequently, we expect that alloy composition may influence the magnitude of the dislocation share.

\section{Summary}
In this work, we have utilized in-situ diffraction and post mortem microscopy in conjunction with dominant shear deformation to identify, dissect, and quantify the relevant deformation mechanisms in nanocrystalline (NC) $\mathrm{Pd}_{90}\mathrm{Au}_{10}$ in the limiting case of $D \leq \unit[10]{nm}$. We could identify lattice and grain boundary (GB) elasticity, shear shuffling or shear transformations (STs) operating in the core region of GBs, stress driven grain boundary migration (SDGBM), and dislocation shear along lattice planes to all contribute, however, with significantly different and non-trivial stress-dependent shares to overall deformation. The frequently used term GB sliding has been discarded here since STs, which act as generic flow event in metallic glasses, have been unambigiously identified to also operate as dominant strain carriers in the core region of GBs \cite{Grewer2014}.

{It was not required to invoke nonlinear elasticity to adequately describe the material response. In the so-called microplastic regime, STs coexist with lattice and GB elasticity in a way that the share of the latter decreased inversely proportional to the progressive increase of STs. Dislocation activity contributes with a share of less than 10\% in this regime, implying that it rather acts as accommodation mechanism. As a matter of fact, STs and elastic contributions by far dominate the material response in this regime, therefore, it seems plausible to discard the conventional concept of work- or strain hardening to account for the evolution of the tangent modulus. This assessment is consistent with a recent finding unveiling that NC Pd-Au alloy-systems manifest solid solution softening at the low end of the nanoscale \cite{Leibner2015} and not dislocation-based hardening behavior.}   

{Moreover, STs which propagate strain at/along GBs have been found to carry about two thirds of the overall strain in the regime of macroplasticity (beyond the yield stress). SDGBM exhibits an appreciable decrease of activity in this regime, whereas dislocation activity progressively increases. Overall, the diverse evolution of the respective deformation shares appears as being just a consequence of a complex hierarchical order of onset/nucleation and upholding stresses of different deformation mechanisms. We expect that the configurational state of the material's microstructure, whether tested in as-prepared or relaxed state, should have a decisive influence on this hierarchy.}

\begin{addendum}
 \item The authors are grateful for the preparation of the FIB lamellae and TEM dark field micrographs taken by Aaron Kobler at the Karlsruhe Micro Nano Facility (KMNF), the synchrotron beamtime provided by the European Synchrotron Radiation Facility (ESRF experiments MA1112 \& MA1353), and financial support from Deutsche Forschungsgemeinschaft (FOR 714 and BI 385/18-1). The authors also want to thank Werner Skrotzki for critically reading the manuscript and providing helpful comments and suggestions, particularly with respect to texture formation. Further thanks go to Andreas Leibner for fruitful discussions, as well as assistance with data reduction.

 \item[Correspondence] Correspondence should be addressed to Christian Braun~(email: c.braun@nano.uni-saarland.de).
\end{addendum}

\bibliographystyle{elsarticle-num} 					
\bibliography{References_Mechanics_of_Materials}



\end{document}